\documentclass[%
reprint,
 amsmath,amssymb,
 aps,
 pre,
]{revtex4-2}

\usepackage{graphicx}
\usepackage{dcolumn}
\usepackage{bm}
\usepackage{comment}
\usepackage{color}
\usepackage[normalem]{ulem}
\newcommand \bl{\color{black}}
\newcommand \rd{\color{black}}

\newcommand \mg{\color{black}}
\begin{document}

\preprint{APS/123-QED}

\title{Squirmer hydrodynamics near a periodic surface topography}

\author{Kenta Ishimoto$^{1}$*, Eamonn A. Gaffney$^2$ and David J. Smith$^3$}
\affiliation{%
$^1$
 Research Institute for Mathematical Sciences, Kyoto University,
 Kyoto 606-8502, Japan \\
$^2$
Wolfson Centre for Mathematical Biology, Mathematical Institute, University of Oxford, Oxford OX2 6GG, UK\\
$^3$
School of Mathematics, University of Birmingham, Birmingham B15 2TT, UK
}%


%

\date{\today}

\begin{abstract}

\vspace*{1mm} 

The behaviour of microscopic swimmers has previously been explored near large scale confining geometries and {\mg in} the presence of very small-scale surface roughness. Here we consider an intermediate case of how a simple microswimmer, the tangential spherical squirmer, behaves adjacent to {\rd  singly- and doubly-periodic sinusoidal surface topographies} that {\mg spatially} oscillate with an amplitude that is an order of magnitude less than the swimmer size and  wavelengths   within an order of magnitude of this scale.   The nearest neighbour regularised Stokeslet method is used for numerical explorations after validating its accuracy for a spherical tangential squirmer that swims stably near a flat surface. The same squirmer is then introduced to {\mg the} different surface topographies. {\mg The  key governing factor in the resulting swimming behaviour is the size of the squirmer relative to the surface topography wavelength. When the squirmer is much larger, or much smaller, than the surface topography  wavelength then directional guidance is not observed. Once the squirmer size is on the scale of the topography wavelength  limited guidance is possible, often with local capture in the topography troughs. However,   complex
dynamics can also emerge, especially when the initial configuration is not close to alignment along topography troughs or above topography crests. In contrast to sensitivity in alignment and the topography wavelength,} reductions  in the amplitude of the surface topography or variations in the shape of the periodic surface topography do not have extensive impacts {\bl on} the squirmer behaviour.  Our findings  more generally highlight   that  detailed  {\mg numerical} explorations  are  necessary to determine how surface topographies may interact with a given  swimmer that swims stably near a flat surface and whether surface topographies   may be designed to provide  such swimmers with directional guidance.

\vspace*{3mm}
\end{abstract}

\maketitle


\section{Introduction}

 
Ever since the studies of  Robert Hooke and Antonie van Leeuwenhoek it has long been known that a drop of pond water contains countless microbes, many of which are motile and indeed the very  occasional one can be  lethal, such as  {\it Naegleria fowleri}, the causative agent of amoebic meningitis. Even harmless motile microbes have attracted the attention of scientists for centuries,  though 
nowadays  developments in nano- and micro-technology has also enabled the fabrication of  self-propelling artificial micro-robots and the  manipulation of their dynamics using microfluidic devices \cite{kherzi2016}. In laboratory experiments and observations,  with both synthetic and biological swimmers,  confinement and control are necessary, though typically  crude and often limited to the inevitable confinements of wall boundaries. 

Nonetheless,   the physical effects of    walls on microswimmers can be subtle and has been extensively investigated theoretically, numerically, and experimentally in the past decade \cite{lauga2009, elgeti2015}. For instance, numerous biological microswimmers such as bacteria and sperm cells are known to accumulate near a flat wall boundary (\cite{kantsler2013, bianchi2017, ohmura2018}). Furthermore, motility near surfaces also has  a functional role, for instance   biofilm formation and initial spread \cite{pratt1998},   
as well as     enhanced searching,  which in turn is significant for fish egg fertilisation, where sperm need to encounter the egg micropyle \cite{cosson2008}.
In addition, curved boundaries such as   convex walls,  corners and obstacles, are easily fabricated in microfluidic devices, which has motivated studies on the effects of such confinements both for biological microorganisms \cite{denissenko2012, tung2014, sipos2015, shum2015, ishimoto2016, nosrati2016, ostapenko2018, nishiguchi2018, rode2019, yang2019} and artificial microswimmers
\cite{takagi2014, spagnolie2015, liu2016, yang2016, wykes2017}.

These curved boundaries and obstacles are typically larger than, or comparable to, the swimmer. If the structure on the wall boundary is smaller than the swimmer lengthscale, it may considered as a rough boundary instead of a completely flat surface. The impact of surface roughness  has previously been  considered via an  asymptotically small amplitude of the surface topography in the presence of a spherical particle   and a spherical microswimmer \cite{rad2010,assoudi2018}, the so-called the squirmer \cite{shaik2017}.


The squirmer is a model microswimmer first proposed by Lighthill \cite{lighthill1952} as a slightly deforming sphere and later used by Blake \cite{blake1971a} as a model of ciliate swimmers.  This simple model is currently understood to provide qualitative predictions for a spherical biological microswimmer \cite{pedley2016, pedley2016a}. In particular, a simplified version of the model, in which a rigid sphere can self-propel due to a given surface velocity slip profile, is known as the spherical tangential squirmer. This has been widely used as a simple mathematical model with a finite volume for studies on hydrodynamical aspects of microswimming such as nutrition uptake \cite{magar2003}, cell-cell interactions \cite{ishikawa2006, drescher2009}, Janus particle motility   \cite{spagnolie2012,ishimoto2013}, collective dynamics \cite{evans2011, zottl2014, delfau2016, oyama2016}, swimming in {\mg a} non-Newtonian medium \cite{lauga2009a, zhu2012, nganguia2018} and swimming near a wall \cite{llopis2010, spagnolie2012, ishimoto2013}.
The squirmer has also been studied in the context of confinement  and obstacles such as the interior of a tube \cite{zhu2013}, the presence of lattice-like multiple obstacles \cite{chamolly2017}, or  a curved and structured wall \cite{das2019}.

Investigations into the effects of small surface topography {\mg on  microswimmers} are, however, limited to the asymptotic analysis of rough surfaces {\bl or}  boundary features, such {\mg as} curvatures {\mg with} lengthscales that are much larger than the swimmer.  
The purpose of the current study is, therefore, to bridge the gap between an asymptotically small amplitude of the surface roughness and large lengthscale curved boundaries. In particular for periodic structures at this mesoscale, there is the prospect that  the microswimmer may become oriented and   guided by the surface, and we will    numerically investigate the dynamics of a spherical tangential squirmer under these conditions. Such investigations are particularly  motivated by recent studies {\mg of}  a colloidal microswimmer near a small surface topography \cite{simmchen2016}, highlighting that {\mg a} structured  surface topography  may be fabricated in a microfluidic devices with the potential for utilisation in the guidance of microswimmers.

The very near-wall dynamics, at a separation of around 50nm and less  \cite{dickinson2003}, typically requires both hydrodynamic and steric interactions \cite{dickinson2003, kantsler2013, bianchi2017}, and a short-range repulsive potential force is often considered for a modelling study \cite{spagnolie2012}. However, even a small difference {\mg in the} repulsion function  can alter swimmer behaviour  \cite{lintuvuori2016, ishimoto2017}. {\mg Thus,} in this initial study, we only focus on the hydrodynamic interactions, and do not consider any additional steric interactions and contact mechanics.

{\mg This is   motivated not only by the utility of relative simplicity in this initial study, but also for understanding the impact of hydrodynamic surface interactions where, despite attractive steric forces, swimmer deposition is  undesirable and thus of minimal interest, compared to topography guidance dynamics when deposition does not occur. It also entails that the results and conclusions of this study are not contingent on the details of contact mechanics and steric forces, which vary with surfaces and solutes \cite{dickinson2003}. Thus, the numerical simulations are stopped just before the squirmer dynamics is influenced by the short-range   dynamics  on very close surface approach. This short-range  detail may be accommodated in later work together   with many further refinements,  such as incorporating more faithful representations of flagellated and ciliated swimmers.}

 
The structure of this paper is as follows. We introduce the squirmer model and three different  surface topographies in Sec. \ref{sec:squirmer}. In Sec. \ref{sec:numer}, our numerical methods and their verifications are discussed. We will then present the simulation results for the different surface topographies in Sec. \ref{sec:results}, followed by a discussion of the implications, in particular  for     microswimmer guidance via surface topography, in  Sec. \ref{sec:disc}.





\section{The squirmer}
\label{sec:squirmer}

We consider the non-dimensional Stokes equations of the low-Reynolds-number flow, from which it follows for a velocity field $\bm{u}$ and a pressure field $p$ that
\begin{equation}
\nabla p= \Delta\bm{u}~~\textrm{and}~~\nabla\cdot\bm{u}=0
\label{eq:M020}.
\end{equation}
We impose the no-slip boundary condition on the swimmer and the wall, together with the force and torque balance equations for {\mg a swimmer with} negligible inertia. 

\begin{figure}
\begin{center}\includegraphics[bb= 0 0 510 355, width=8cm]{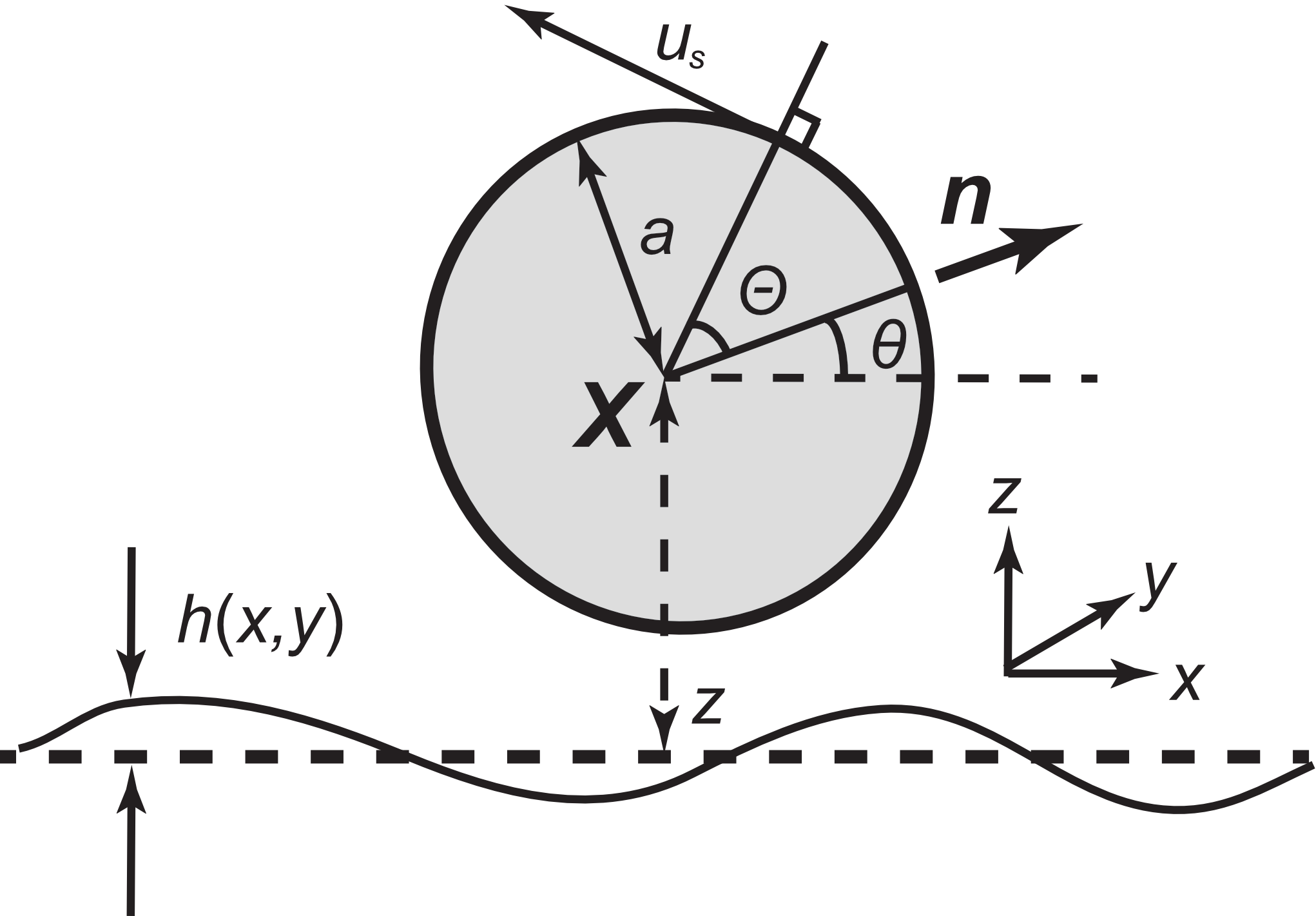}
\caption{Schematic diagram of a   spherical tangential squirmer, of radius $a=1$,  near a no-slip wall. Here, ${\bf X}$ denotes the centre of the spherical swimmer relative to a laboratory reference frame, with Cartesian coordinates $(x,y,z)$ and $h(x,y)$ is the height of the surface above its average midplane (dashed) at $z=0$. Further, $z$ is overloaded and also represents the height of the swimmer centre above the midplane, with an analogous overloading of $x$, $y$. Whether $x,y,z$ refer to coordinates or the overloaded definition $\bm{X}=(x,y,z)$ for the location of the squirmer centre,  will be clear from context. The unit vector $\bm{n}$ gives the orientation of the swimmer, which makes an angle $\theta$  relative to the midplane of the surface topography and $\Theta$ is the local polar angle of a point on the swimmer surface  relative to $\bm{n}$. The swimmer's motility is driven by an axisymmetric  tangential  velocity of its surface, of size  $u_s$, and in the direction of increasing $\Theta$, as detailed in the main text. 
}
\label{fig:conf}
\end{center}
\end{figure}

We first introduce the  spherical tangential squirmer model. We have a spherical swimmer 
of radius $a=1$, possessing 
an axisymmetric and tangential surface velocity (e.g., \cite{ishikawa2006}). The sphere  centre is located at $\bm{X}=(x,y,z)$, and  $\bm{n}$ denotes the  unit vector of its orientation, as shown schematically in Fig. \ref{fig:conf}, where coordinates,   variables for the position of the squirmer centre,  and the angles $\theta,~\Theta$ are defined with aid of a {\mg diagram.}  
In particular,  $\Theta\in[0,\pi]$ denotes the polar angle relative to ${\bf n}$, and we  impose an axisymmetric tangential velocity slip $u_s$ on the squirmer, given by 
\begin{equation}
u_s(\Theta)=\sum_{n=1}^{\infty}B_nV_n(\cos\Theta)
\label{eq:M10a},
\end{equation}
where  $V_n$ is a function derived from the Legendre polynomial $P_n(x)$ via
\begin{equation}
V_n(x)=\frac{2\sqrt{1-x^2}}{n(n+1)}\frac{d P_n(x)}{dx}
\label{eq:M10b}.
\end{equation}
The swimming velocity in free space is dictated by the first mode, with $\bm{U}=(2/3)B_1\bm{n}$ \cite{lighthill1952,blake1971a}. Throughout this paper, we fix $B_1=3/2$ so that the squirmer swimming speed is set to be unity in free space and  
 we neglect the higher modes, by setting $B_n=0$ for $n\geq 3$
so that the swimmer is subsequently fully  characterised  by the   squirmer parameter   $\beta=B_2/B_1$ \cite{ishikawa2006}.
 In particular, and following convention,  the swimmer is denoted as  a pusher when $\beta<0$, a puller when $\beta>0$ and  a neutral swimmer for $\beta=0$ \cite{evans2011}.  
 The second mode, associated with the parameter $B_2$,  corresponds to the flow {\mg induced} by the Stokes dipole.  In particular, a cell with a trailing flagellum, such as an {\it E. Coli} bacterium or a sperm cell, behaves as a pusher, while   cells with leading flagella, such as {\it Chlamydomonas} and {\it Leishmania} \cite{walker2019}, are modelled as   pullers, with cells possessing {\bl fore}-aft symmetry, such as ciliates, behaving as neutral swimmers \cite{evans2011}.

\begin{figure}
\begin{center}
\includegraphics[bb=0 0 1399 472,  width=9cm]{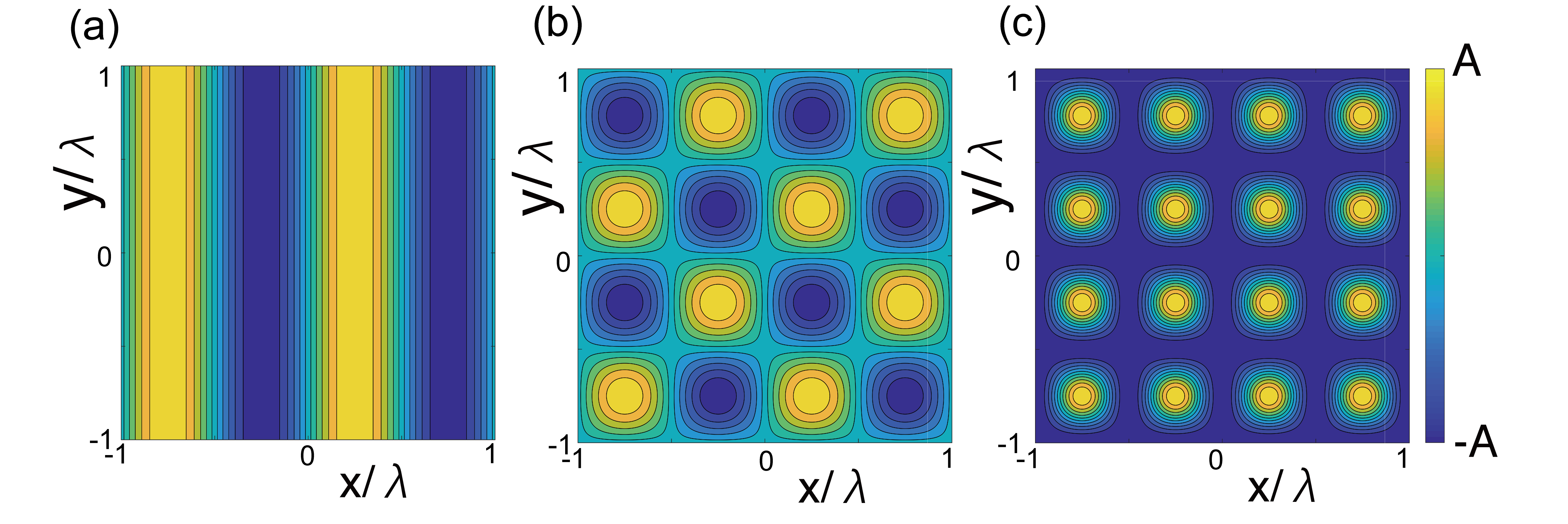}
\caption{Illustrations of the surface topography functions $h(x,y)$ considered in this study. (a) A singly-periodic sinusoidal wave topography, (b) a doubly-periodic sinusoidal wave and  (c) a doubly-periodic peaks, respectively associated with Eqns. (\ref{eq:w1}), (\ref{eq:w2}), (\ref{eq:w3}. The maximum and minimum heights, $h(x,y)$, are $+A$ and $-A$ for all topographies.}
\label{fig:wall}
\end{center}
\end{figure}

Here, we focus on spherical tangential  squirmers that swim stably near a flat  surface and thus we consider puller squirmers with   ($\beta\gtrsim 5$), which  are known to {\mg exhibit  stable} swimming near a flat surface \cite{ishimoto2013,uspal2015}.
In particular, we examine their dynamics close to   surfaces with  structured topographies. The first of these topographies is a one-dimensional  sinusoid defined by    
\begin{equation}
h(x,y)=A\sin(kx)
\label{eq:w1},
\end{equation}
where $A$ is the   amplitude and $k=2\pi/\lambda$ is the wavenumber, with $\lambda$ denoting the wavelength (Fig. \ref{fig:wall}a). The second topography is given by the doubly-periodic sinusoid    
\begin{equation}
h(x,y)=A\sin(kx)\sin(ky)
\label{eq:w2}, 
\end{equation}
as depicted in (Fig. \ref{fig:wall}b), and the third is given by 
\begin{equation}
h(x,y)=A\left[ 2\sin^2(kx)\sin^2(ky)-1\right]
\label{eq:w3}.
\end{equation}
This final topography  is shown in   Fig. \ref{fig:wall}c and presents doubly-periodic peaks  with    highest and lowest heights of  $+A$ and $-A$, respectively, as in the previous two topographies. However, note the inter-peak wavelength is now halved  and the parameter $\lambda =2\pi/k$ no longer represents the wavelength since the sinusoidal functions are squared in Eqn. (\ref{eq:w3}). Throughout this study for the doubly periodic topographies, we focus on cases  that are symmetrical in switching the $x$ and $y$ directions. 

We consider both the surface of the  squirmer and the wall surface topography, which we denote by $S$ and $W$, respectively, with the boundary conditions of the Stokes equations given by  no-slip conditions on both boundaries. The surface velocity of the squirmer can  be described by the combinations of the squirmer linear velocity $\bm{U}$ and   angular velocity $\bm{\Omega}$, together with its tangential surface   velocity, $\bm{u}_s$, of size given by Eqn. (\ref{eq:M10a}) and in the axisymmetric tangential direction, as depicted in Fig. \ref{fig:conf}. Hence, the no slip condition entails the fluid velocity on the surface of the swimmer is given by 
\begin{equation}
\bm{v}(\bm{x})=\bm{U}+\bm{\Omega}\times(\bm{x}-\bm{X})+\bm{u}_s,~~
(\bm{x}\in\,S)
\label{eq:M01a}.
\end{equation}
In contrast, the wall surface topography is assumed to be stationary, and we thus have
\begin{equation}
\bm{v}(\bm{x})=\bm{0} ~~(\bm{x}\in W)
\label{eq:M01b}.
\end{equation}

\section{Numerical methods}
\label{sec:numer}

\subsection{Nearest neighbour regularised Stokeslet method}

The dynamics of the squirmer has been computed, using the nearest-neighbour regularised Stokeslet method (nnRSM) \cite{smith2018, gallagher2018}, and the numerical simulations have been  performed based on the Matlab code accompanied by \cite{gallagher2018}, as we now summarise. The Stokes flow boundary integral equations for the  single-layer formulation are  given by  \cite{pozrikidis1992} 
\begin{equation}
u_j(\bm{x})=-\frac{1}{8\pi}\int_{{\bl S\cup W}} S_{ij}(\bm{x},\bm{y})f_i(\bm{y})\,dS_{\bm{y}}.
\end{equation}
Here, $f_i(\bm{y})$ denotes the components of the surface traction and the integral kernel $S_{ij}$ is the Stokeslet, which exhibits an integrable singularity as $\bm{y}\rightarrow\bm{x}$, with  the surface integral   well-defined. For  numerical tractability Cortez introduced a regularised Stokeslet \cite{cortez2001}, {\rd which is the exact divergence-free solution to the Stokes flow equations with a spatially-smoothed point force, {\mg  and then}}  approximated the boundary integral \cite{cortez2005} via 
\begin{equation}
u_j(\bm{x})=-\frac{1}{8\pi}\int_{{\bl S\cup W}} S^{\epsilon}_{ij}(\bm{x},\bm{y})f_i(\bm{y})\,dS_{\bm{y}}
\label{eq:M01c},
\end{equation}
where $S^{\epsilon}_{ij}$ is the regularised Stokeslet and $\epsilon$ is the regularisation parameter. We recover the singular boundary integral, once we take the limit of $\epsilon\rightarrow 0$. Following \cite{cortez2005}, we consider a regularised Stokeslet of the form,
\begin{equation}
S^{\epsilon}_{ij}(\bm{x},\bm{y})=\left\{\left(r^2+2\epsilon^2\right)\delta_{ij}+r_ir_j\right\}\left(r^2+\epsilon^2\right)^{-3/2},
\end{equation}
where $\bm{r}=\bm{x}-\bm{y}$, $r=|\bm{r}|$, and $\delta_{ij}$ is Kronecker's delta. The no-slip boundary conditions are simply given by enforcing  $\bm{u}(\bm{x})=\bm{v}(\bm{x})$ from Eqns. (\ref{eq:M01a},\ref{eq:M01b}) for boundary points in the integral equation (\ref{eq:M01c}). Since the squirmer is swimming freely, we also have the inertialess force and torque balance equations,
\begin{equation}
\int_{S}\bm{f}(\bm{x})\,dS_{\bm{x}}=\int_{S}(\bm{x}-\bm{X})\times\bm{f}(\bm{x})\,dS_{\bm{x}}=\bm{0}
\label{eq:M02}.
\end{equation}

We then discretise the surface integrals (\ref{eq:M01c}), (\ref{eq:M02}), by introducing the quadrature node  positions $\bm{x}[n]$ and the associated weights $A[n]$ for the discretised surface point $n$ ($n=1,\cdots,N$), where $N$ is the total number of the surface points. The above surface integrals contain the product of `$\bm{f}\,dS$' and we discretise the integral \cite{smith2018,gallagher2018} via
\begin{equation}
\int \bullet\, f_j(\bm{x})\,dS_{\bm{x}} \approx   \sum_{n=1}^N \bullet\, g_j[n]A[n]  
\label{eq:M03},
\end{equation}
where {\mg on the righthand side} the symbol, $\bullet$, represents an arbitrary function of $\bm{x}$, {\rd  evaluated at $\bm{x}[n]$}, and $g_j[n]=f_j(\bm{x}[n])$. 

\begin{figure}
\begin{center}
\includegraphics[bb=0 0 1105 963, width=7cm]{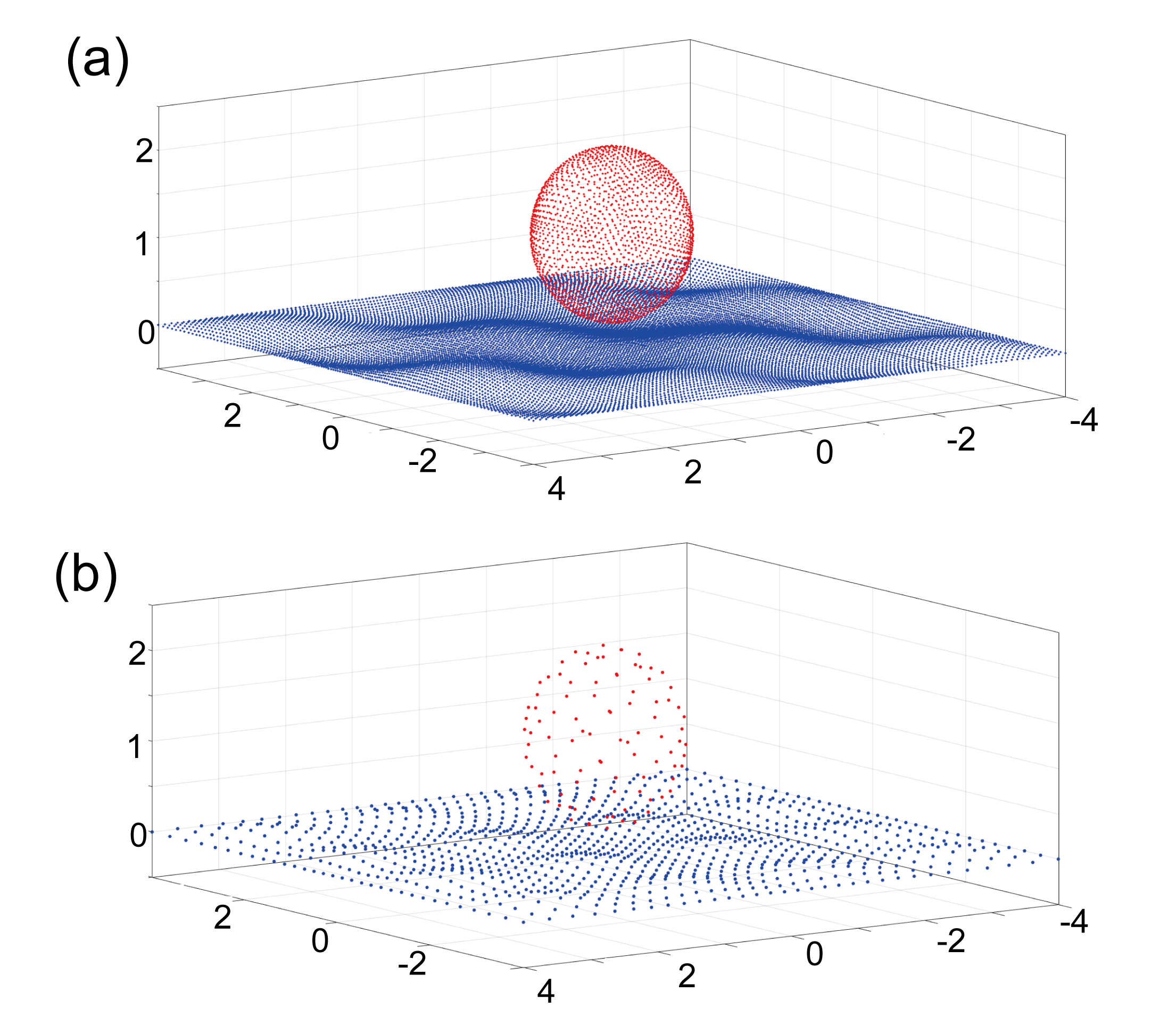}
\vspace{1em}
\caption{An illustration  of the points representing the squirmer and   surface topography with the finer   discretisation (a) used for the kernel  and the  coarser  discretisation (b),   used for the surface traction. Discretisation points on the  spherical squirmer surface are indicated by red dots. In contrast   blue discretisation points are shown for the doubly periodic surface topography of Fig \ref{fig:wall}b, 
with the size of the discretised surface given by   $L=8$, with the surface topography given by Eqn.~(\ref{eq:w2}) and plotted in Fig.~\ref{fig:wall}b, with  an     amplitude of $A=0.1$ and the wavelength given by  $\lambda=2$.  }
\label{fig:mesh}
\end{center}
\end{figure}

{\mg Continuing} with framework of Smith \cite{smith2018}, we introduce a second  surface discretisation, $\bm{x}[q]$,  $(q=1,\cdots,Q)$ which corresponds to a more refined   discretisation than used for the surface traction, with $N\ll Q$, as illustrated in Fig.~\ref{fig:mesh}. The two discretisations enable efficient numerical solution as the kernel, $S_{ij}^\epsilon$, can vary rapidly and thus requires a finer discretisation, which would be inefficient if used for the surface traction, $\bm{f}$. 
{\rd Moreover, if the force and quadrature discretisations {\mg do not} overlap, the quadrature error no longer diverges as $\epsilon \rightarrow 0$, {\mg and hence  a less refined force discretisation in this framework} is in general more accurate than if the two discretisations coincide.}

  The nearest-neighbour matrix, $\nu[q,n]$, is then defined {\rd separately for a swimmer and a wall} as 
\begin{equation}
\begin{cases}
1 & \textrm{if} ~ n=\arg\min|\bm{x}[\hat{n}]-\bm{x}[q]| ~ \left( \bm{x}[{\rd \hat{n}}], \bm{x}[q]\in S\right) \\
1 & \textrm{if} ~~ n=\arg\min|\bm{x}[\hat{n}]-\bm{x}[q]|
~ \left(\bm{x}[{\rd \hat{n}}], \bm{x}[q]\in  W\right) \\
0 & \textrm{otherwise},
\end{cases}
\label{eq:M04}
\end{equation}
where $\arg\min$ is the argument at which the minimum is attained over the set $\hat{n}\in\{1,\cdots,N\}$ and we use this   matrix to interpolate the discretisation via 
\begin{equation}
f_j(\bm{x}[q])dS(\bm{x}[q])\approx\sum_{n=1}^N \nu[q,n]g_i[n]A[n]
\label{eq:M05}.
\end{equation}

Combining Eqns. (\ref{eq:M03},\ref{eq:M05}) and noting the total number of the points for the finer discretisation is $Q$, we have
\begin{equation}
\int \bullet(\bm{x})\, f_j(\bm{x})\,dS_{\bm{x}} \approx
\sum_{n=1}^N g_j[n]A[n]
\sum_{q=1}^Q \bullet(\bm{x}[q])\,\nu[q,n]
\label{eq:M06}.
\end{equation}
We use $\bullet(\bm{x})=S_{ij}^\epsilon(\bm{x}',\bm{x})$ in Eqn.~(\ref{eq:M06})  to discretise the boundary integral (\ref{eq:M01c}), and we use $\bullet(\bm{x})=1$ and $\bullet(\bm{x})=\epsilon_{ijk}(x_k-X_k)$ in Eqn.~(\ref{eq:M03}) for the force and torque balance relations (\ref{eq:M02}), respectively.

In particular to solve for the squirmer trajectory, we first need to determine its velocity, $\bm{U}$, and angular velocity $\bm{\Omega}$, which can then be integrated over time to determine the squirmer location and orientation. Firstly note that at a fixed point in the time,  the squirmer location and orientation are known from previous integration or from the initial conditions at the start of the simulation.  Then the discretisations of Eqns.~(\ref{eq:M01c},\ref{eq:M02}),   with $\bm{u}=\bm{v}$ eliminated in terms of $\bm{U},$ $ \bm{\Omega}$ and the known $\bm{u}_s$ via Eqns.~(\ref{eq:M10a},~\ref{eq:M10b},~\ref{eq:M01a},~\ref{eq:M01b}), gives $3N+6$ constraints for the $3N+6$ scalar unknowns associated with the unknown surface tractions at the $N$
discretisation points and the unknowns $\bm{U},$ $ \bm{\Omega}$. The resulting linear system is readily solved. 

As is the case for both singular and regularised versions of the boundary integral representation for flow around a constant volume body, the integral equation admits a gauge freedom $\bm{f}\rightarrow \bm{f}+\alpha\bm{m}$, where $\alpha$ is any constant and $\bm{m}$ is the surface normal pointing into the fluid (this can be observed by applying incompressibility and the divergence theorem). In the absence of boundary conditions for traction, this freedom results in the pressure being determined only up to an additive constant. Discretisation results in an invertible matrix and hence a unique (approximate) solution; moreover the non-uniqueness {\mg of the continuum  solution for the pressure} is not dynamically important as it does not affect either the total force or moment on the swimmer.

\subsection{Swimming in a free space}

We first examine  the numerical accuracy of the swimming velocity calculation for the squirmer in {\it free space.} The squirmer parameter is set to  $\beta=0$, and the exact swimming speed   is  $|\bm{U}|=1$, as detailed in the previous section. We have fixed the {\bl regularisation} parameter $\epsilon=0.001$, and have examined the impact of changing  the discretisation refinement. In particular, with   the total number of the points that form the squirmer surface given by $N=6n_s^2$ and $Q=6N_s^2$ for each discretisation \cite{smith2018, gallagher2018}, changes in both $n_s$ and $N_s$ have been examined. These are as summarised in Table~\ref{tab:param} where an  accuracy of approximately $1\%$ is observed, though   finer meshes do not   improve  this accuracy. However, for a fixed regularisation parameter, this is often reported in methods using regularised Stokeslets (e.g. \cite{cortez2005}).   Finally we also note 
that changes in $\beta$ have not been observed to  alter the swimming speed, as expected. 

\begin{table}[tb]
\begin{center}
\caption{Predictions for the free space swimmer speed. The exact speed is given by $|\bm{U}|=1$ and its numerical calculation is presented for refinements of both the discretisations used in the nearest neighbour regularised Stokeslet method, where $N=6n_s^2$ and $Q=6N_s^2$}  
\vspace*{4mm} 
\begin{tabular}{cccc}
 \hline  ~$\epsilon$~ & ~$n_s$~& ~$N_s$~ & ~$|\bm{U}|$~ \\ \hline
0.001 & 4 & 10 &1.0135 \\
0.001 & 4 & 12 &1.0017 \\
0.001 & 4 & 14 &1.0073 \\
0.001 & 4 & 16 &1.0153 \\
0.001 & 4 & 18 &1.0048 \\ \hline
0.001 & 5 & 10 &1.0291 \\
0.001 & 5 & 12 &1.0079 \\
0.001 & 5 & 14 &1.0098 \\
0.001 & 5 & 16 &1.0113 \\ \hline
\end{tabular}
\label{tab:param}  
\end{center}
\end{table}
\setlength{\parskip}{5pt}


\subsection{Swimming near a wall}

Hereafter, we  set   the squirmer discretisation parameters to be  $(n_s, N_s)=(4, 18)$ and 
we  proceed to consider the squirmer near a no-slip wall. As previously studied by the boundary element method \cite{ishimoto2013}, a strong puller tends to  {\mg stably} swim near a {\it flat} wall. We therefore choose $\beta=7$, and   set the initial location of the squirmer centre to be  $(0,0,1.15)$, with  the  initial orientation given by  $\theta=-0.017\pi$, which is effectively the initial angle of attack relative to the midplane of the surface topography, as can be seen from Fig.~\ref{fig:conf}. 

\begin{figure}
\begin{center}
\includegraphics[bb= 0 0 737 517, width=6.5cm]{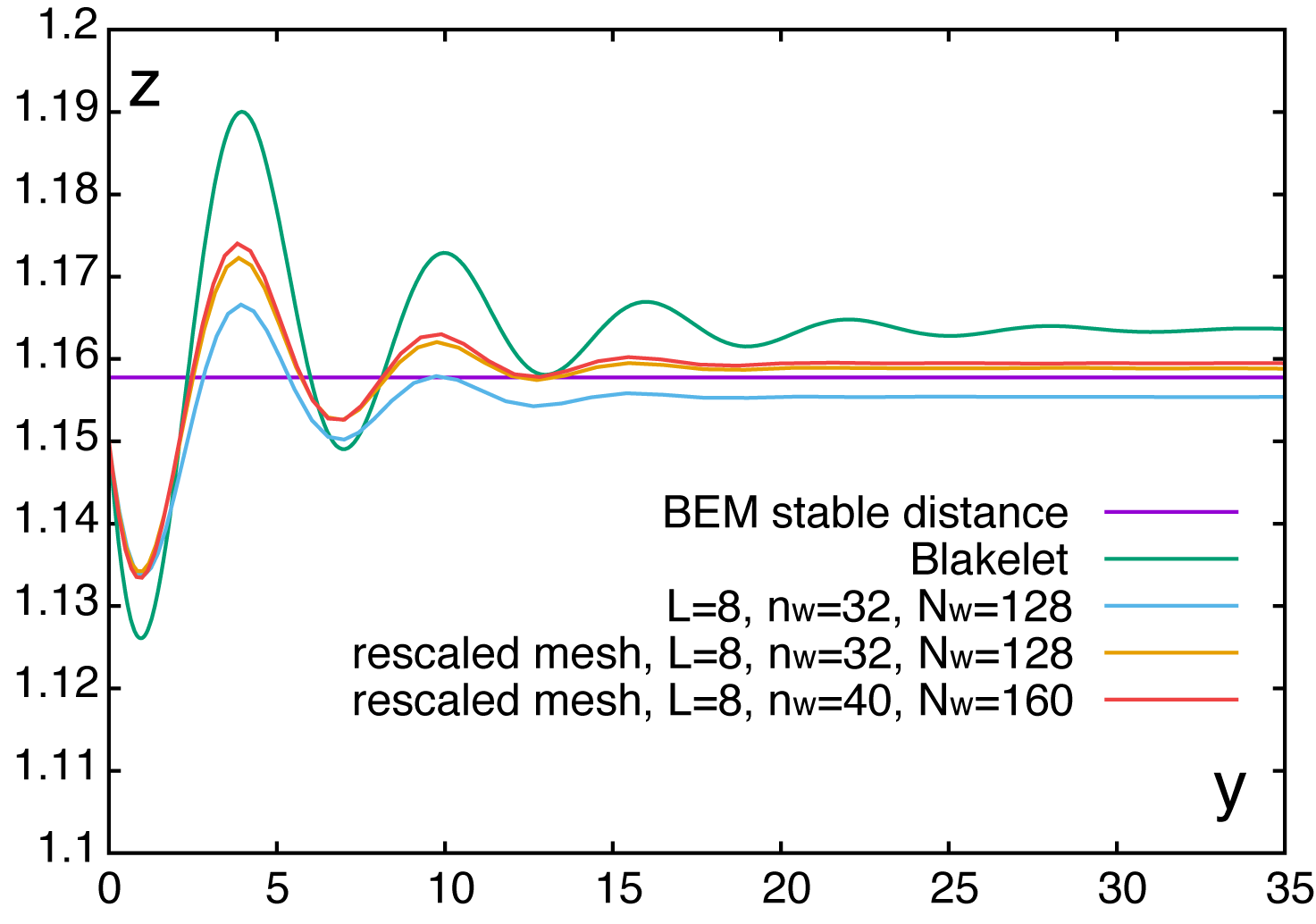}
\vspace{1em}
\caption{Swimming trajectories of the squirmer with a regularisation of $\epsilon=0.001$ and different discretisation parameters. Also plotted is the height of the stable fixed point obtained by the boundary element method using the singular Blakelet \cite{ishimoto2013}, as labelled by ``BEM stable distance'', and a trajectory using a regularised Blakelet with the nearest neighbour regularised Stokeslet method, as labelled by ``Blakelet''.  }
\label{fig:flat}
\end{center}
\end{figure}

We first use the regularised Blakelet \cite{ainley2008} in the nearest neighbour regularised Stokeslet method  and compare the simulation result with the stable distance obtained via the boundary element method using the singular Blakelet \cite{ishimoto2013}, with the latter providing the stable separation distance $z^\ast\approx1.1578$. As shown in Figure \ref{fig:flat}, these two algorithms' predictions are in reasonable agreement.

We then consider a wall that is captured by an explicit discretisation of its surface rather than by use of the Blakelet,    as implemented in the sperm simulation of Gallagher et~al.~\cite{gallagher2018}. The wall is given by the $x$-$y$ plane and  represented by the square with a length of $L$, with its centre     at the location of the projection of the squirmer centre, $\bm{X}$, onto the plane $z=0$.
Each side contains $n_w$ and $N_w$ points with equal separations for respective surface traction  and kernel discretisations \cite{smith2018, gallagher2018}. Hence the number of  points  on  the surface  $S\cup W$ are given by $N=6n_s^2+n_w^2$ and $Q=6N_s^2+N_w^2$ for the surface traction  and kernel discretisations, respectively. An  example {\mg of a} swimming trajectory is  plotted in Figure \ref{fig:flat}.

We then rescale the wall points to resolve the squirmer-boundary hydrodynamic interaction more efficiently by using the function 
$$f: [-1/2, 1/2] \rightarrow [-1/2, 1/2], 
~~~~
f(x)=\frac{1}{2} \tan{\left( \frac{\pi x}{2} \right)}.
$$
The equally discretised square of unit length $$ {\cal S}= \{ (x,y) \in [-1/2,1/2] \times[-1/2,1/2] \}$$ is mapped by this function, via $$\{ (f(x),f(y)) ; (x,y) \in {\cal S} \},$$ and then dilated by the scale of $L$. The square obtained by this scheme allows a more precise representation of   the hydrodynamical interactions between the squirmer and the wall,  as seen from the results   labelled by ``rescaled" in  {\mg Fig. \ref{fig:flat},} which use this mapping. These trajectories in particular are sufficiently accurate for our purposes and very close    to the prediction of the   boundary element method (BEM)  prediction of the stable swimming height above the surface, which is exact to within discretisation error. 


\section{Results}
\label{sec:results}

\begin{figure}
\begin{center}
\includegraphics[bb=0 0 538 511,  width=6cm]{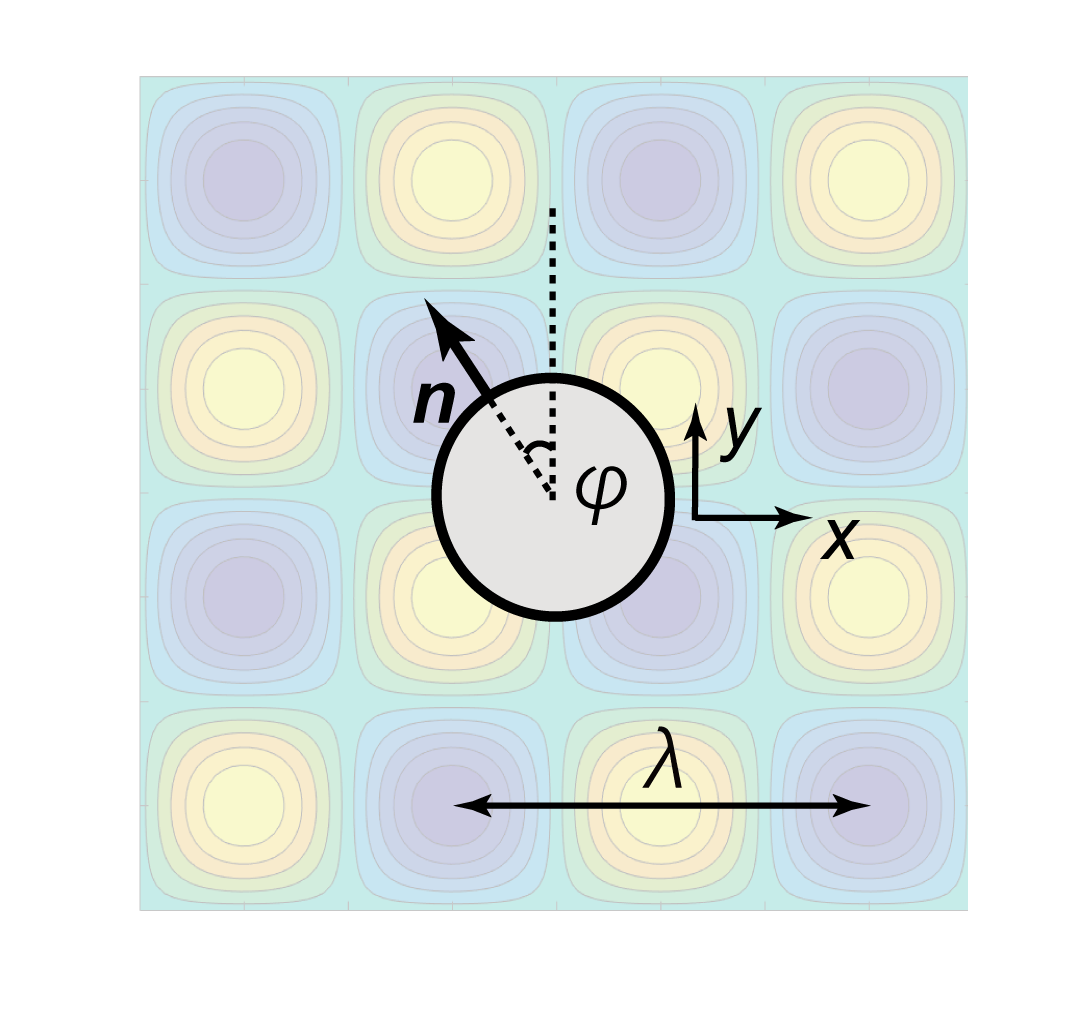}
\caption{A bird's eye view  of the squirmer above the doubly periodic surface topography of Eqn.~(\ref{eq:w2}), as depicted in Fig. \ref{fig:wall}b, with $\lambda=4$. The angle $\varphi$ is defined to be  angle of the $x$-$y$ projection of the squirmer orientation vector $\bm{n}$, from the $y$-axis, as illustrated, and thus gives the direction of the squirmer in the $x$-$y$ plane.  }
\label{fig:conf2}
\end{center}
\end{figure}

In this section, we discuss the swimming trajectories of the squirmer near a surface with a structured periodic topography, as defined in Eqns. (\ref{eq:w1})-(\ref{eq:w3}) and depicted in Fig. \ref{fig:wall}. For all simulations presented below the initial height of the squirmer was fixed at $z=1.2$, with the initial angle of attack given by $\theta=-0.17\pi$. Further 
 initial squirmer centre location coordinates of  $x=0$,  $y=0$, are set together with  squirmer parameters of $B_1=3/2,~\beta=7$, and a surface topography amplitude of $A=0.1$, unless explicitly stated otherwise. The surface topography wavelength and the initial orientation of the squirmer in the $x$-$y$ plane, namely $\varphi$ in Fig.~\ref{fig:conf2}, are varied extensively among the simulations and  either stated or, in the case of $\varphi$,  can otherwise  be immediately inferred from the initial tangent angle of the trajectories in the presented plots.

\subsection{Singly-periodic sinusoidal topography}

We start with the single wave sinusoid topography of  Eqn.~(\ref{eq:w1}) and Fig. \ref{fig:wall}a. The initial location of the squirmer and the initial angle of attack are fixed at the simulation default initial values of $\bm{X}=(0,0,1.2)$ and $\theta = -0.17\pi$, as stated  above,     while the initial orientation of the squirmer in the $x$-$y$ plane  has been initially considered in detail with $\varphi=0.5\pi$ and then subsequently varied.

\begin{figure}
\begin{center}\includegraphics[bb= 0 0 869 694, width=8.5cm]{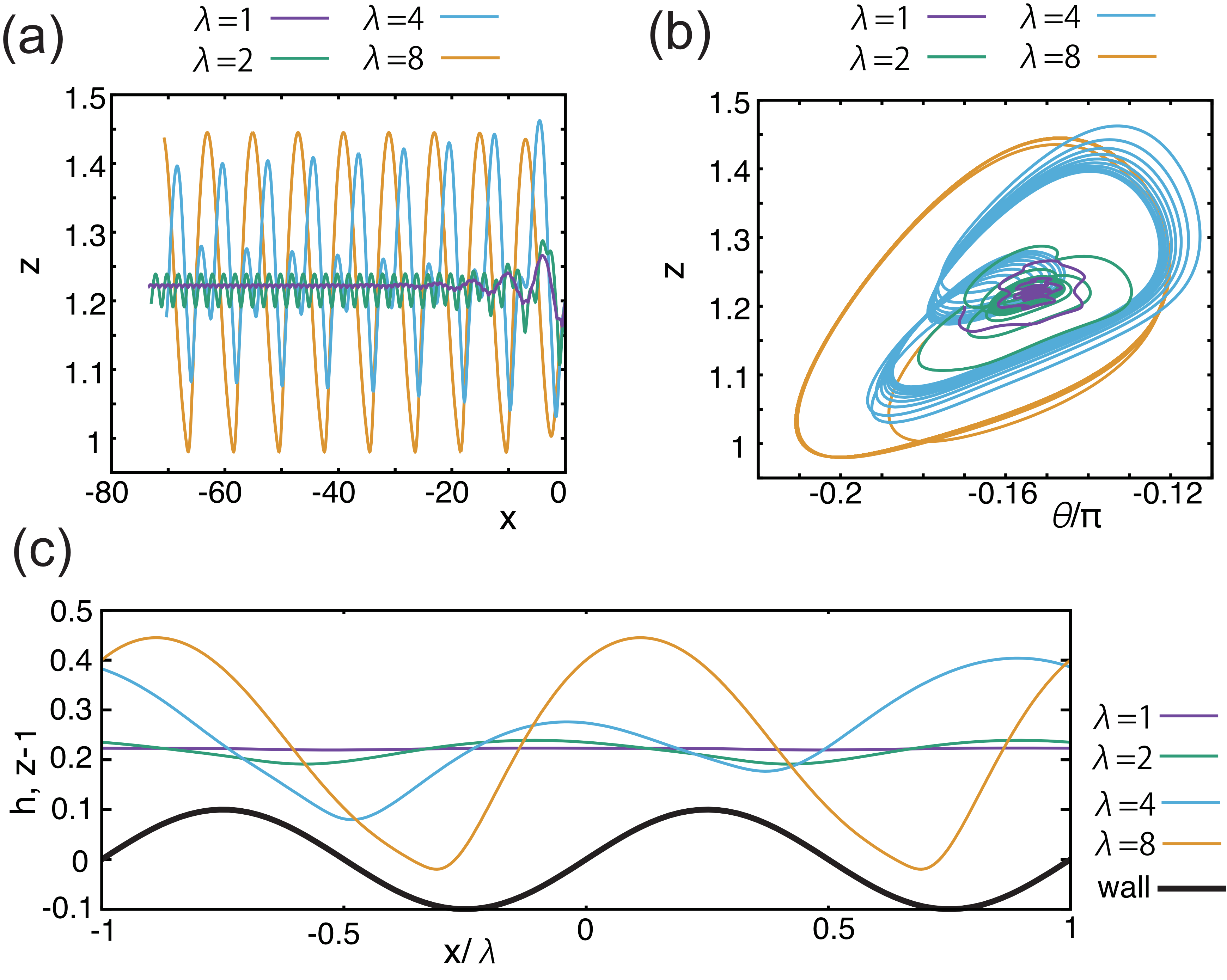}
\caption{Dynamics of the squirmer   swimming adjacent to the  singly-periodic sinusoidal topography of Eqn.~(\ref{eq:w1}) and Fig. \ref{fig:wall}a, with different wavelengths $\lambda=1, 2, 4, 8$. The wave amplitude of the topography is fixed at $A=0.1$. (a) Trajectories in the $x$-$z$ plane. (b) Trajectories in the $\theta$-$z$ phase plane. (c) Horizontally rescaled and shifted $z$ position for the last part of the simulation (a), together with the topography function $h$. 
}
\label{fig:phi0c1}
\end{center}
\end{figure}

Thus, we first consider a squirmer swimming parallel to the wave vector of the sinusoidal topography, or equivalently along the $x$-axis. Fixing the initial orientation relative to the $x$-$y$ plane via $\varphi=0.5\pi$, swimming trajectories in the $x$-$z$ plane are plotted in Fig.~\ref{fig:phi0c1}a for different surface topography wavelengths $\lambda=1, 2, 4, 8$. Corresponding trajectories in the $\theta$-$z$ phase plane are shown in Fig.~\ref{fig:phi0c1}b. When the wavenumber $\lambda$ is smaller ($\lambda=1, 2$), the squirmer attains stable oscillatory swimming, but the oscillation in the $z$-axis is smaller than the surface topography amplitude, $A=0.1$, highlighting  that the topography only perturbs the stable position associated with swimming near a flat wall. This may be additionally observed in Fig.~\ref{fig:phi0c1}c, where the $z$ dynamics for the last part of the oscillating motion obtained in \ref{fig:phi0c1}a are shown relative to a  horizontally rescaled and shifted axis $x/\lambda$ together with the surface topography function $h$.  However, as the wavelength is increased to $\lambda=4,$ $\lambda=8$, the oscillatory motion then transitions to   an  amplitude that is larger than that of the surface topography, as can be seen in  Fig. \ref{fig:phi0c1}c. In addition one can observe in this figure, with $\lambda=4$, that the wavelength of the $z$-component  oscillations in the trajectory need not match that of the underlying surface topography, though these two wavelengths do match once the surface topography wavelength has been increased to $\lambda=8.$

\begin{figure*}
\begin{center}
\includegraphics[bb= 0 0 900 713, width=14cm]{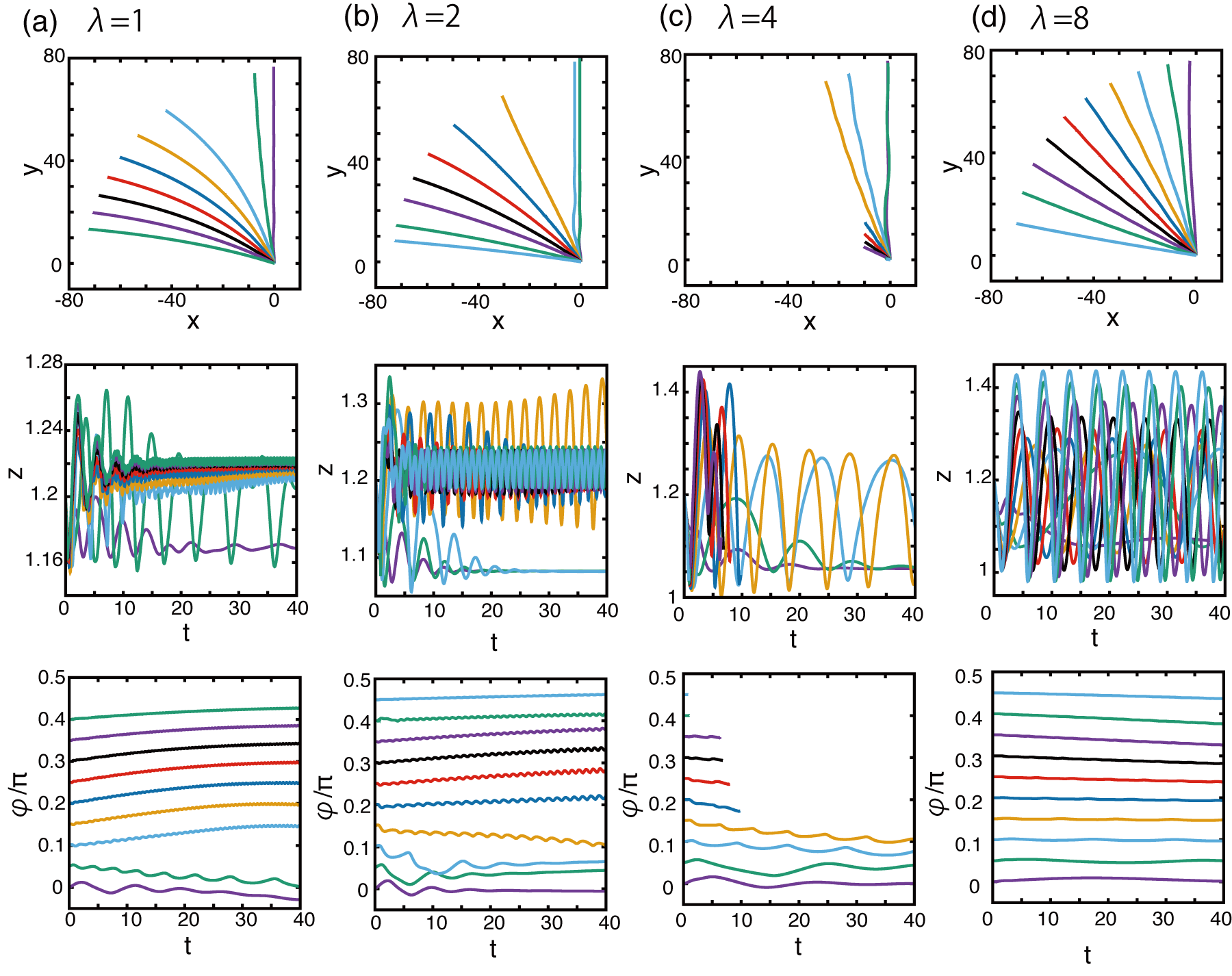}
\vspace{1em}
\caption{Dynamics of the squirmer near a surface with the singly-periodic sinusoidal wave topography of Eqn.~(\ref{eq:w1}) and Fig. \ref{fig:wall}a. 
The surface topography amplitude is given by   $A=0.1$  and the wavelength is (a) $\lambda=1$, (b) $\lambda=2$, and (c) $\lambda=4$, (d) $\lambda=8$. (Top panels) The projections of the squirmer trajectories   onto the $x$-$y$ plane with different initial orientation angles, $\varphi$, as defined via Fig.~\ref{fig:conf2}. These initial angles may be inferred from the initial tangents of the plotted projected trajectories.      (Middle panels) The time evolution of the height of the centre of the squirmer, $z$. (Bottom panels) The time evolution of the orientation angle $\varphi$. Different colours index   different initial values of the orientation angle, $\varphi$.}
\label{fig:w1}
\end{center}
\end{figure*}

We then proceed to consider the squirmer dynamics with   different initial  values of $\varphi$, and thus different initial orientations relative to the $x$-$y$ plane, while once more varying the wavelength of the singly periodic topography. 
In Fig. \ref{fig:w1}, we plot the predicted trajectories and the orientations for the case with the surface topography amplitude $A=0.1$ and  wavelength $\lambda=1, 2, 4, 8$, while considering various values of $\varphi$, from zero to $0.5\pi$.
From the figure, one can observe  that the squirmer tends to swim either parallel or perpendicular to the direction of the well, which is aligned along the  $y$-axis, though the orientation angle $\varphi$ need not always necessarily match  the  direction of motion. Hence, an overall  drifting motion can be induced by the squirmer-topography hydrodynamic interaction.

Notably, the swimming dynamics associated with an initial   orientation angle of  $\varphi\approx 0.05$ with $\lambda=1$, or {\bl $\varphi\approx 0.15$} with $\lambda=2$,  is unstable and the trajectories evolve towards {\bl  the stable orientations of $\varphi=0$}, as seen in Fig. \ref{fig:w1}a,b. Here, the stable swimming along the $y$-axis  is accompanied by  hydrodynamic capturing, with the squirmer moving  along a trough of the surface topography. Furthermore, in both cases, the $z$ dynamics attains a stable oscillatory motion, as may be observed in Fig. \ref{fig:phi0c1}a,b. 
In both cases, the qualitative aspects of these features are unaltered when the amplitude $A$ is {\mg decreased,} though the timescale for the reorientation along one of the axes becomes longer. 

Analogously,  an increase in the topography wavelength to the intermediate value    of   $\lambda=4$ entails that swimming oblique to the troughs and peaks of  topography can be observed, as seen in  
Fig.~\ref{fig:w1}(c). However, at this wavelength 
the squirmer concomitantly undergoes extensive oscillations in the $z$ direction. Furthermore, the squirmer  enters the near vicinity of the surface 
(Fig.~\ref{fig:w1}(c)) once it is no longer oriented approximately along the $y$-axis. {\mg This is at the stage of requiring the consideration   of  surface  mechanics, which is outside the detailed scope of the study and hence we stop the numerical simulation, thus also avoiding the numerically unreasonable spatial resolutions required for the   associated fine scale  hydrodynamics.}


We further increase the wavelength of the sinusoidal topography to $\lambda=8$, with trajectories
presented in Fig.~\ref{fig:w1}d, which on projection to the  $x$-$y$ plane are essentially straight,  regardless of the initial orientation angle $\varphi$. Hence, at larger wavelengths,   the squirmer swims    with a direction that is unaffected by the surface. Furthermore, the $z$-dynamics of the squirmer trajectory  become more oscillatory as the topography wavelength is increased, as observed previously  in Fig. \ref{fig:phi0c1}, unless the squirmer   is captured in the trough  along the $y$-axis, in which case the $z$ dynamics converges to a   stable position. 

More generally, all of these observations highlight that even with a surface topography amplitude of $A=0.1$, which to the eye is small as highlighted by  Fig.~\ref{fig:mesh},   the squirmer behaviour is affected by the structured surface topography in a complex manner. In particular, the resulting  trajectories are contingent on the details of the  topography parameters and the squirmer orientation, especially once the  topography  wavelength is comparable to the squirmer size.

\subsection{Doubly-periodic sinusoidal topography}

We now consider  the squirmer  dynamics near a surface with  the doubly-periodic wave topography, given by Eqn.  (\ref{eq:w2}) and illustrated in Fig.~\ref{fig:wall}b.  In the current setting, the  topography  breaks the translational symmetry in the $y$-direction, and hence the trajectories now also depend on the value of the initial $y$ position of the squirmer centre. 
We first consider the squirmer starting  with orientation  $\varphi=0.5\pi$ and an initial centre location coordinate of  $y=\lambda/4$, with the other initial location coordinates  and the initial  angle of attack  at the default values of $x=0$, $z=1.2$ and $\theta=-0.17\pi$. Then the squirmer    moves  along the $-x$ axis with  similar  dynamics in the $z$-direction
to that displayed in Fig.~\ref{fig:phi0c1}. 
In particular, the 
 dynamics in the $z$-direction are only slightly perturbed when $\lambda=1, 2$ but  display larger oscillations when  $\lambda=4, 8$. 

\begin{figure*}
\begin{center}
\includegraphics[bb=0 0 935 467, width=13cm]{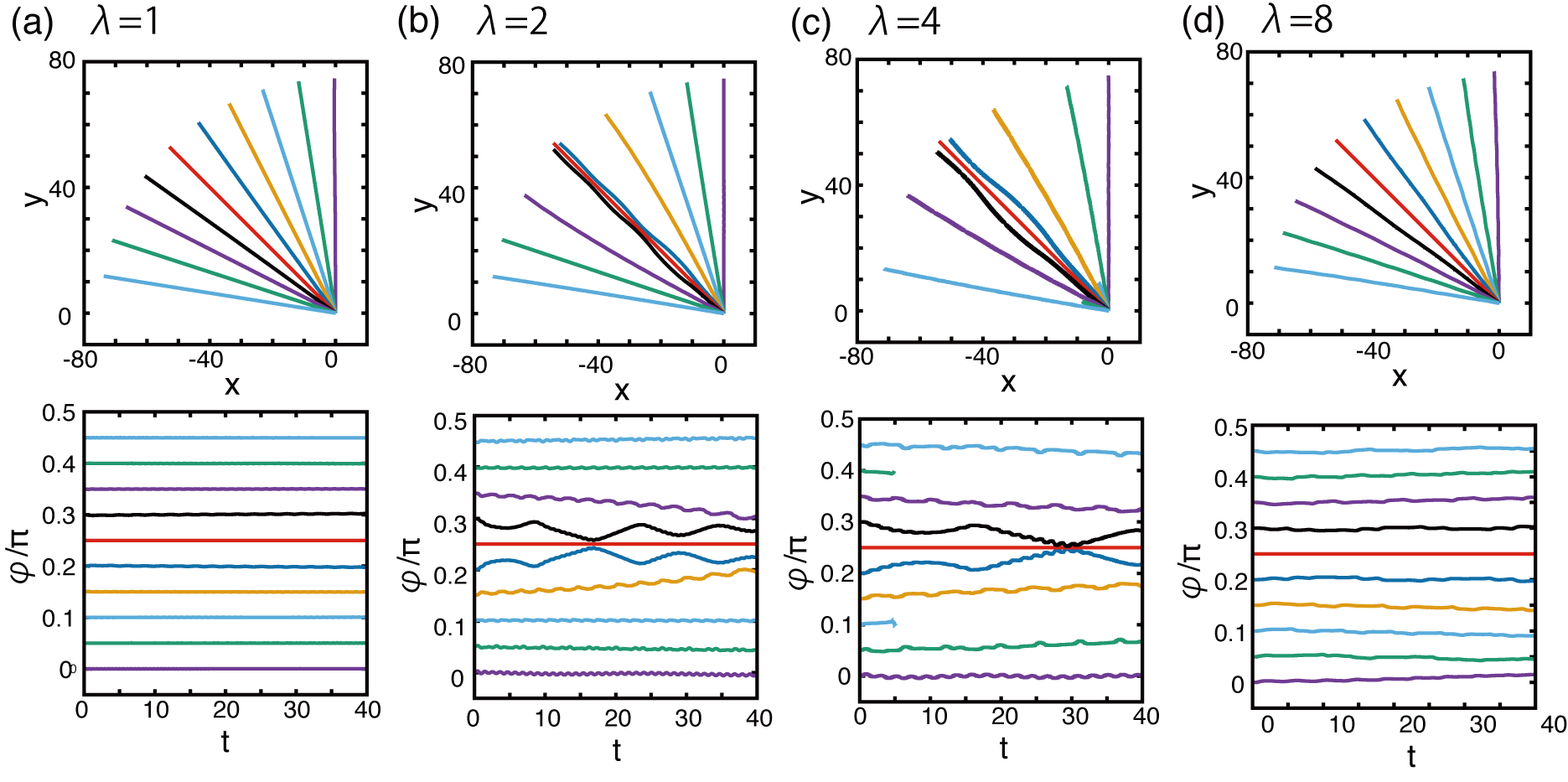}
\vspace{1em}
\caption{Dynamics of the squirmer near a surface with the doubly-periodic sinusoidal wave topography of Eqn. (\ref{eq:w2}), as depicted in Fig.~\ref{fig:wall}b. The surface topography amplitude is given by $A=0.1$  and the wavelength is (a) $\lambda=1$,  (b) $\lambda=2$, (c) $\lambda=4$, {\mg (d)} $\lambda=8$. (Top panels) The projections of the squirmer trajectories   onto the $x$-$y$ plane with different initial orientation angles, $\varphi$, as defined via Fig.~\ref{fig:conf2}. These initial angles may be inferred from the initial tangents of the plotted projected trajectories.    (Bottom panels) The time evolution of the orientation angle $\varphi$. Different colours index    different initial conditions.}
\label{fig:w2}
\end{center}
\end{figure*}

We then proceed to consider variations in the initial 
squirmer orientation angle $\varphi$ that, as previously, entails the trajectories are no longer constrained to two spatial dimensions. The surface topography amplitude remains  fixed at  $A=0.1$ and the initial height, location and attack angle are at default values, while we consider  variation in the orientation angle  $\varphi$ and the topography wavelength, which here is still given by  $\lambda$. When $\lambda=1$ (Fig. \ref{fig:w2}a), the trajectories are not affected by the surface topography, as the trajectories are straight when projected on to the $x$-$y$ plane, and the angle $\varphi$ is constant in time.
However, in the simulation with the wavelength $\lambda=2$, some trajectories with $\varphi\approx \pi/4$ are hydrodynamically captured near the bottom of the doubly-periodic sinusoidal valley, whereas  swimming outside of this  region of initial orientation angles is  not affected by the surface topography, as in the case of $\lambda=1$.    

\begin{figure*}
\begin{center}
\includegraphics[bb= 0 0 895 485, width=13cm]{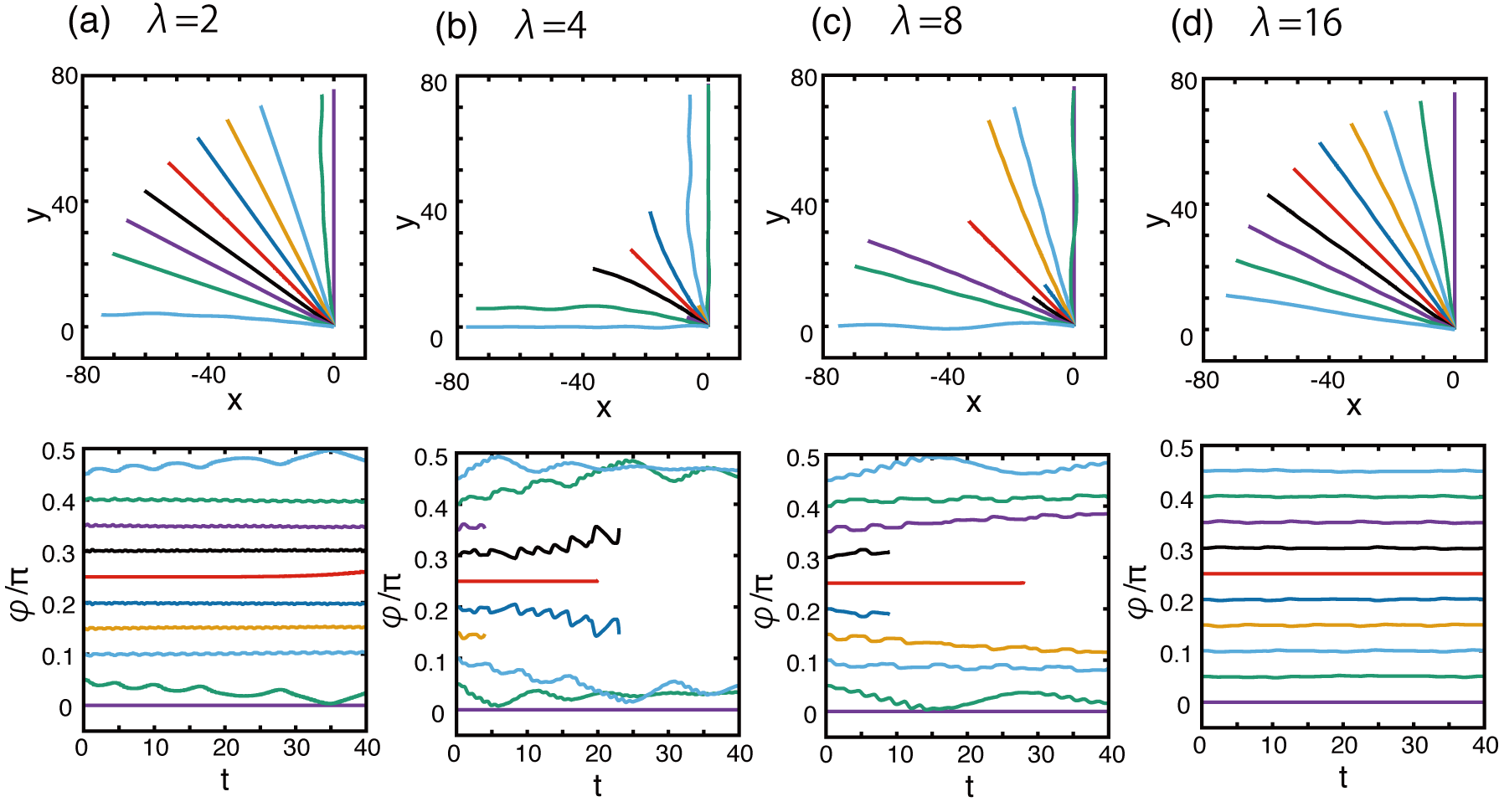}
\vspace{1em}
\caption{Dynamics of the squirmer near a surface with the doubly-periodic sinusoidal wave topography of Eqn. (\ref{eq:w3}), as depicted in Fig.~\ref{fig:wall}c. The surface topography amplitude is given by $A=0.1$  and the wavelength is unchanged from previous plots, but is no longer given by $\lambda=2\pi/k$ where $k$ is the wavenumber of the 
sinusoidal functions in Eqn. (\ref{eq:w3}), since these functions are  squared. Hence, to preserve wavelength at $1,2,4,8$ length units  in the respective columns, we take 
(a)~$\lambda= 2$,  (b)~$\lambda=4$, (c)~$\lambda=8$, 
(d)~$\lambda=16$ on moving from left to right across the figure.  (Top panels) The projections of the squirmer trajectories   onto the $x$-$y$ plane with different initial orientation angles, $\varphi$, as defined via Fig.~\ref{fig:conf2}.  These initial angles may be inferred from the initial tangents of the plotted projected trajectories. (Bottom panels) The time evolution of the orientation angle $\varphi$. Different colours index    different initial conditions.}
\label{fig:w3}
\end{center}
\end{figure*}

These features can also be observed  when we increase the wavelength to $\lambda=4$ (Fig. \ref{fig:w2}c). In contrast, with the larger wavelength of $\lambda=8$,   there is no evidence for an attracting basin of  squirmer dynamics near the bottom of the topographic valley (Fig. \ref{fig:w2}d). Together these results highlight  that the hydrodynamic attraction toward, and subsequently along, topographic valleys is not only limited, but also only possible when the scale of the swimmer diameter is comparable with the lengthscale of the surface topography.

We then move to consider the final surface topography of  doubly-periodic peaks as introduced by Eq. (\ref{eq:w3}) and displayed in Fig. \ref{fig:wall}c. 
Trajectories with straight line projections onto the $x$-$y$ plane can be observed when the   squirmer   initial location and orientation angle 
{\mg align along topography troughs or across topography crests.}
 For example, given the default initial location of  $\bm{X}=(0,0,1.2)$ and an initial orientation angle $\varphi=0$,   the squirmer swims with $\varphi=0$ throughout time. In addition, these trajectories also exhibit nearly constant $z$-dynamics, though small $z$-oscillations  are observed with amplitude $\lesssim 0.05$, due to the topography and this oscillation is further diminished as the topography wavelength increases. 
Furthermore, with the default initial location and an initial value of  $\varphi=0.25\pi$, or with   initial values of $\bm{X}=(0,\lambda/4,1.2)$ and $\varphi=0.5\pi$, straight line $x$-$y$ projected trajectories are also observed. Furthermore, in the $z$-direction the squirmer behaviour changes from small oscillatory perturbations   to larger amplitude $z$-oscillations as the wavelength increases, in direct analogy  to examples considered in detail with the previous topographies.

For more general initial configurations of the squirmer, we have considered the three-dimensional behaviour of the resulting trajectories,  with the simulation results   plotted in Fig. \ref{fig:w3} for    $\lambda=2, 4, 8, 16$, noting the wavelength for this topography is $\lambda/2$, not $\lambda$. In the previous doubly periodic surface topography,  swimming with the orientation  angle $\varphi=\pi/4$ allowed the cell to move along  a surface topography trough, noting the prospect of drifting  observed in the singly periodic topography of Fig.~\ref{fig:wall}a can be ruled out. In contrast, for the current case, an orientation of $\varphi=\pi/4$   moves across the surface topography peaks.

Once more, the simulation results  show that the squirmer is attracted by orientation  angles corresponding to troughs of the surface topography, which here are $\varphi=0,~\pi/2$ for example. Furthermore, such topographic attraction is realised when the squirmer diameter is close to the characteristic lengthscale of the surface modulation, as seen in Fig.~\ref{fig:w3}a,b,c. However, once more for the  parameter regimes of Fig.~\ref{fig:w3}b,c
the  swimmer can  approach  very close to the 
surface   
and the trajectory simulations are halted as {\rd detailed surface dynamics} are outside the scope of the study. Finally, we note that once the surface topography wavelength is increased further, with $\lambda=16$ as presented in  Fig.~\ref{fig:w3}d, all projected trajectories are straight lines and the orientation angle $\varphi$ is constant in time. Hence, at these larger wavelengths  the surface-swimmer interactions   no longer influence the guidance of the squirmer.

\section{Discussion}
\label{sec:disc}

In this paper, we have numerically investigated the hydrodynamics of a puller spherical tangential squirmer near a surface with a singly- or doubly-periodic structured topography. In particular, the surface topography possessed an amplitude an order of magnitude less than the squirmer size and a wavelength on the scale of the squirmer size. The  simulated squirmer is known to be attracted to a stable separation  from a flat wall, and a mesh-free regularised Stokeslet boundary element numerical scheme was demonstrated to accurately  capture the   dynamics induced  by the subtle hydrodynamic interactions between a spherical tangential squirmer and a flat wall.

When a wavelength of the sinusoidal surface topography is smaller than the squirmer size, the perpendicular dynamics of the swimmer trajectory  is a small amplitude oscillatory  perturbation  from {\mg the constant stable swimming height associated with a flat boundary.} However as the wavelength of the surface topography is increased,  the squirmer acquires larger vertical,  $z$-direction, oscillations with a wavelength that matches that of the topography at very large surface topography wavelengths but not always at intermediate values (Fig.~\ref{fig:phi0c1}c).

Furthermore, the squirmer movement in the horizontal, $x$-$y$,  plane   has been observed to be highly dependent on the detailed geometrical properties of the surface topography. 
We first considered a singly-periodic sinusoidal surface topography.  When the wavelength of the surface  topography (Eqn.~\ref{eq:w1})  is not significantly larger than the squirmer diameter  ($\lambda~<~4$),
  the  horizontal squirmer motion reorientates  towards the nearer of   two stable {\mg directions,} which are parallel and perpendicular to the wavevector of the surface topography, as seen in Fig.~\ref{fig:w1}. Furthermore, drifting can sometimes be observed, whereby the direction of motion differs slightly from the orientation angle, $\varphi.$
  
At intermediate wavelengths ($\lambda=4)$ the  squirmer 
can  approach   
the surface. The detailed subsequent behaviour would be contingent on the {\mg  near-surface dynamics, the detailed study of which  is} beyond the   scope of this study. Once the surface topography wavelength is further increased, with $\lambda=8$ sufficient so that the wavelength is four times that of the squirmer diameter, we see the horizontal motion is that of straight lines and surface induced guidance of the squirmer in the horizontal plane is lost. 
 
For the doubly-periodic  surface topographies,   the squirmer had the tendency  {\mg to be locally guided to} swim along  surface topography troughs rather than over crests, though  only when the squirmer diameter was  comparable to the surface topography  wavelength.   However, such tendencies were weak and, at intermediate surface topography lengthscales, often  accompanied by very close approaches to the surface where   steric interactions would be important.


These qualitative features of the squirmer trajectories have  also been  observed, when the squirmer parameter, $\beta=B_2/B_1$ from Eqn.~\ref{eq:M10a}, was varied within the range  that induced  {\mg stable swimming} in the vicinity of a flat wall. In addition, changing   the sinusoidal surface topography to a sigmoidal topography to represent fabricated surface wells in microfluidic devices, did not induce a significant change in the qualitative features of the swimming trajectories. In turn, this evidences that   squirmer  swimming behaviours are influenced mainly by the lengthscale of the surface topography.
{\mg In addition, we have also observed that the swimmer behaviour can be complex, especially when the swimmer is not aligned along the surface topography troughs or above the surface topography crests, though a local guidance to swim along surface topography troughs has also been common.}


There are considerable numbers of studies focusing on microswimmer dynamics near a non-trivial geometrical structures such as curved obstacles \cite{nishiguchi2018, das2019}, bumps \cite{simmchen2016, yang2019} and maze-like micro-devices \cite{denissenko2012, tung2014}, but the lengthscale of the surface topography  in the current study features much finer surface structures. We also note  that the transitional vertical behaviours in the $z$-direction,   from perturbations of the stable swimming height for a flat wall  to topography-following motion at very large surface topography wavelength, necessitates a consideration of a finite-size amplitude of the surface  topography. In particular,  such behaviours highlight that the dynamics examined in this study requires larger scale physics beyond the effective boundary conditions \cite{sarkar1996,kunert2010,luchini2013} based on a very small amplitude surface roughness.

In the context of representing  biological microswimmers such as spermatozoa  and bacteria, which are pusher swimmers rather than pullers,   hydrodynamic stable swimming occurs for   prolate pusher tangential squirmers, but not   spherical squirmers \cite{ishimoto2013}. Moreover, the hydrodynamic interactions strongly depends on the swimmer morphology and beating pattern \cite{ishimoto2014, walker2019}. In turn, this highlights that  detailed numerical studies are required in further work to explore the surface dynamics for both  prolate squirmers and  more realistic microbiological swimmers near   non-trivial  surface topographies, {\mg including the prospect of  a  ciliated epithelium, modelled    as a dynamic periodic boundary \cite{smith2008}.}

Further, contact dynamics are also experimentally known to be significant for boundary accumulation behaviours of microswimmers \cite{kantsler2013, bianchi2017} and to vary extensively with    solutes and surfaces \cite{dickinson2003}. However, the current study does not consider the {\rd detailed surface dynamics}   in the region very close {\mg to} the boundary, since its scope considers universal hydrodynamic interactions, rather than the boundary behaviours for a specific swimmer, solute and surface. Even with a simple short-range repulsion, the details of the   repulsive force can alter the  swimmer  dynamics \cite{lintuvuori2016, ishimoto2017}, while  the   contact mechanics      reflect the specific  biological and  physical features of the system under investigation.  
A further generalisation that should be considered in the context of artificial colloidal microswimmers is the impact of sedimentation and gyrotaxis  due to a swimmer  density  heterogeneity  and a density offset from the fluid \cite{das2020}, as well the chemical and physical   mechanisms that drive the colloidal particle \cite{uspal2015}

{\mg In summary, this investigation  has used the nearest neighbour regularised boundary element method \cite{gallagher2020} to numerically explore the   hydrodynamic interactions between a   spherical tangential squirmer and a spatially  oscillating surface topography, 
  with amplitude that is an order of magnitude less than the squirmer size.
 In particular 
a squirmer was investigated  that 
swam with very simple dynamics close to a flat boundary, relaxing to a stable distance from the wall and swimming in the direction of its orientation in the horizontal plane.
However, even with small amplitude surface topographies, this squirmer's dynamics has depended in a subtle and complex manner on  the 
wavelength of the surface topography. We 
 found that surface topographies could effect a limited and local squirmer guidance  towards topography troughs, in particular   once the squirmer size}
is of the same order of magnitude as the surface topography wavelength.  However, contact dynamics may also be induced at {\mg such}   wavelengths of the surface topography, {\mg especially  if the initial squirmer orientation to the surface is not along topography crests or troughs.} Nonetheless, 
 surface guided behaviours are robust to other aspects of the surface topography, such as reductions in  the amplitude and changes in the shape of the surface topography waves.  More generally, these observations suggest detailed numerical simulation {\mg is} required for predicting {\mg whether   directional} guidance can be effected by surface topographies for the  many micromachines and microorganisms that surface accumulate.



\vspace*{0.5cm}
\begin{acknowledgments}
K.I. acknowledges JSPS-KAKENHI (18K13456), JSPS Overseas Research Fellowship (29-0146), Kyoto University Hakubi Project, Kyoto University Supporting Program for Interaction-Based Initiative Team Studies (SPIRITS),  MEXT Leading Initiative for Excellent Young Researchers (LEADER) and JST, PRESTO Grant Number JPMJPR1921, Japan. Elements of the numerical simulations were performed within the cluster computer system at the Research Institute for Mathematical Sciences (RIMS) and Institute for Information Management and Communication (IIMC), Kyoto University. The authors are grateful to Dr T. Montenegro-Johnson and Dr M. T. Gallagher for fruitful discussions. 
\end{acknowledgments}

\bibliography{ref}

\providecommand{\noopsort}[1]{}\providecommand{\singleletter}[1]{#1}%
\begin{thebibliography}{65}%
\makeatletter
\providecommand \@ifxundefined [1]{%
 \@ifx{#1\undefined}
}%
\providecommand \@ifnum [1]{%
 \ifnum #1\expandafter \@firstoftwo
 \else \expandafter \@secondoftwo
 \fi
}%
\providecommand \@ifx [1]{%
 \ifx #1\expandafter \@firstoftwo
 \else \expandafter \@secondoftwo
 \fi
}%
\providecommand \natexlab [1]{#1}%
\providecommand \enquote  [1]{``#1''}%
\providecommand \bibnamefont  [1]{#1}%
\providecommand \bibfnamefont [1]{#1}%
\providecommand \citenamefont [1]{#1}%
\providecommand \href@noop [0]{\@secondoftwo}%
\providecommand \href [0]{\begingroup \@sanitize@url \@href}%
\providecommand \@href[1]{\@@startlink{#1}\@@href}%
\providecommand \@@href[1]{\endgroup#1\@@endlink}%
\providecommand \@sanitize@url [0]{\catcode `\\12\catcode `\$12\catcode
  `\&12\catcode `\#12\catcode `\^12\catcode `\_12\catcode `\%12\relax}%
\providecommand \@@startlink[1]{}%
\providecommand \@@endlink[0]{}%
\providecommand \url  [0]{\begingroup\@sanitize@url \@url }%
\providecommand \@url [1]{\endgroup\@href {#1}{\urlprefix }}%
\providecommand \urlprefix  [0]{URL }%
\providecommand \Eprint [0]{\href }%
\providecommand \doibase [0]{https://doi.org/}%
\providecommand \selectlanguage [0]{\@gobble}%
\providecommand \bibinfo  [0]{\@secondoftwo}%
\providecommand \bibfield  [0]{\@secondoftwo}%
\providecommand \translation [1]{[#1]}%
\providecommand \BibitemOpen [0]{}%
\providecommand \bibitemStop [0]{}%
\providecommand \bibitemNoStop [0]{.\EOS\space}%
\providecommand \EOS [0]{\spacefactor3000\relax}%
\providecommand \BibitemShut  [1]{\csname bibitem#1\endcsname}%
\let\auto@bib@innerbib\@empty
\bibitem [{\citenamefont {{B. Kherzi and M. Pumera}}(2016)}]{kherzi2016}%
  \BibitemOpen
  \bibfield  {author} {\bibinfo {author} {\bibnamefont {{B. Kherzi and M.
  Pumera}}},\ }\bibfield  {title} {\bibinfo {title} {Self-propelled autonomous
  nanomotors meet microfluidics},\ }\href@noop {} {\bibfield  {journal}
  {\bibinfo  {journal} {Nanoscale}\ }\textbf {\bibinfo {volume} {8}},\ \bibinfo
  {pages} {17415} (\bibinfo {year} {2016})}\BibitemShut {NoStop}%
\bibitem [{\citenamefont {{E. Lauga and T. R. Powers}}(2009)}]{lauga2009}%
  \BibitemOpen
  \bibfield  {author} {\bibinfo {author} {\bibnamefont {{E. Lauga and T. R.
  Powers}}},\ }\bibfield  {title} {\bibinfo {title} {The hydrodynamics of
  swimming microorganisms},\ }\href@noop {} {\bibfield  {journal} {\bibinfo
  {journal} {Rep.\ Prog.\ Phys.}\ }\textbf {\bibinfo {volume} {72}},\ \bibinfo
  {pages} {096601} (\bibinfo {year} {2009})}\BibitemShut {NoStop}%
\bibitem [{\citenamefont {Elgeti}\ \emph {et~al.}(2015)\citenamefont {Elgeti},
  \citenamefont {Winkler},\ and\ \citenamefont {Gompper}}]{elgeti2015}%
  \BibitemOpen
  \bibfield  {author} {\bibinfo {author} {\bibfnamefont {J.}~\bibnamefont
  {Elgeti}}, \bibinfo {author} {\bibfnamefont {R.~G.}\ \bibnamefont
  {Winkler}},\ and\ \bibinfo {author} {\bibfnamefont {G.}~\bibnamefont
  {Gompper}},\ }\bibfield  {title} {\bibinfo {title} {Physics of
  microswimmers―single particle motion and collective behavior: a review},\
  }\href@noop {} {\bibfield  {journal} {\bibinfo  {journal} {Rep.\ Prog.\
  Phys.}\ }\textbf {\bibinfo {volume} {78}},\ \bibinfo {pages} {056601}
  (\bibinfo {year} {2015})}\BibitemShut {NoStop}%
\bibitem [{\citenamefont {Kantsler}\ \emph {et~al.}(2013)\citenamefont
  {Kantsler}, \citenamefont {Dunkel}, \citenamefont {Polin},\ and\
  \citenamefont {Goldstein}}]{kantsler2013}%
  \BibitemOpen
  \bibfield  {author} {\bibinfo {author} {\bibfnamefont {V.}~\bibnamefont
  {Kantsler}}, \bibinfo {author} {\bibfnamefont {J.}~\bibnamefont {Dunkel}},
  \bibinfo {author} {\bibfnamefont {M.}~\bibnamefont {Polin}},\ and\ \bibinfo
  {author} {\bibfnamefont {R.~E.}\ \bibnamefont {Goldstein}},\ }\bibfield
  {title} {\bibinfo {title} {Ciliary contact interactions dominate surface
  scattering of swimming eukaryotes},\ }\href@noop {} {\bibfield  {journal}
  {\bibinfo  {journal} {Proc.\ Natl.\ Acad.\ Sci.\ USA}\ }\textbf {\bibinfo
  {volume} {110}},\ \bibinfo {pages} {1187} (\bibinfo {year}
  {2013})}\BibitemShut {NoStop}%
\bibitem [{\citenamefont {Bianchi}\ \emph {et~al.}(2017)\citenamefont
  {Bianchi}, \citenamefont {Saglimbeni},\ and\ \citenamefont
  {Leonardo}}]{bianchi2017}%
  \BibitemOpen
  \bibfield  {author} {\bibinfo {author} {\bibfnamefont {S.}~\bibnamefont
  {Bianchi}}, \bibinfo {author} {\bibfnamefont {F.}~\bibnamefont
  {Saglimbeni}},\ and\ \bibinfo {author} {\bibfnamefont {R.~D.}\ \bibnamefont
  {Leonardo}},\ }\bibfield  {title} {\bibinfo {title} {Holographic imaging
  reveals the mechanisms of wall entrapment in swimming bacteria},\ }\href@noop
  {} {\bibfield  {journal} {\bibinfo  {journal} {Phys.\ Rev.\ X}\ }\textbf
  {\bibinfo {volume} {7}},\ \bibinfo {pages} {011010} (\bibinfo {year}
  {2017})}\BibitemShut {NoStop}%
\bibitem [{\citenamefont {Ohmura}\ \emph {et~al.}(2018)\citenamefont {Ohmura},
  \citenamefont {Nishigami}, \citenamefont {Taniguchi}, \citenamefont {Nonaka},
  \citenamefont {Manabe}, \citenamefont {Ishikawa},\ and\ \citenamefont
  {Ichikawa}}]{ohmura2018}%
  \BibitemOpen
  \bibfield  {author} {\bibinfo {author} {\bibfnamefont {T.}~\bibnamefont
  {Ohmura}}, \bibinfo {author} {\bibfnamefont {Y.}~\bibnamefont {Nishigami}},
  \bibinfo {author} {\bibfnamefont {A.}~\bibnamefont {Taniguchi}}, \bibinfo
  {author} {\bibfnamefont {S.}~\bibnamefont {Nonaka}}, \bibinfo {author}
  {\bibfnamefont {J.}~\bibnamefont {Manabe}}, \bibinfo {author} {\bibfnamefont
  {T.}~\bibnamefont {Ishikawa}},\ and\ \bibinfo {author} {\bibfnamefont
  {M.}~\bibnamefont {Ichikawa}},\ }\bibfield  {title} {\bibinfo {title} {Simple
  mechanosense and response of cilia motion reveal the intrinsic habits of
  ciliates},\ }\href@noop {} {\bibfield  {journal} {\bibinfo  {journal} {Proc.\
  Nat.\ Acad.\ Sci.\ U.\ S.\ A.}\ }\textbf {\bibinfo {volume} {115}},\ \bibinfo
  {pages} {3231} (\bibinfo {year} {2018})}\BibitemShut {NoStop}%
\bibitem [{\citenamefont {Pratt}\ and\ \citenamefont
  {Kolter}(1998)}]{pratt1998}%
  \BibitemOpen
  \bibfield  {author} {\bibinfo {author} {\bibfnamefont {L.}~\bibnamefont
  {Pratt}}\ and\ \bibinfo {author} {\bibfnamefont {R.}~\bibnamefont {Kolter}},\
  }\bibfield  {title} {\bibinfo {title} {Genetic analysis of
  $\mbox{E}$scherichia coli biofilm formation: roles of flagella, motility,
  chemotaxis and type $\mbox{I}$ pili},\ }\href@noop {} {\bibfield  {journal}
  {\bibinfo  {journal} {Molecular Microbiology}\ }\textbf {\bibinfo {volume}
  {30}},\ \bibinfo {pages} {285} (\bibinfo {year} {1998})}\BibitemShut
  {NoStop}%
\bibitem [{\citenamefont {Cosson}\ \emph {et~al.}(2008)\citenamefont {Cosson},
  \citenamefont {Groison}, \citenamefont {Suquet}, \citenamefont {Fauvel},
  \citenamefont {Dreanno},\ and\ \citenamefont {Billard}}]{cosson2008}%
  \BibitemOpen
  \bibfield  {author} {\bibinfo {author} {\bibfnamefont {J.}~\bibnamefont
  {Cosson}}, \bibinfo {author} {\bibfnamefont {A.-L.}\ \bibnamefont {Groison}},
  \bibinfo {author} {\bibfnamefont {M.}~\bibnamefont {Suquet}}, \bibinfo
  {author} {\bibfnamefont {C.}~\bibnamefont {Fauvel}}, \bibinfo {author}
  {\bibfnamefont {C.}~\bibnamefont {Dreanno}},\ and\ \bibinfo {author}
  {\bibfnamefont {R.}~\bibnamefont {Billard}},\ }\bibfield  {title} {\bibinfo
  {title} {Marine fish spermatozoa: racing ephemeral swimmers},\ }\href@noop {}
  {\bibfield  {journal} {\bibinfo  {journal} {Reproduction}\ }\textbf {\bibinfo
  {volume} {136}},\ \bibinfo {pages} {277} (\bibinfo {year}
  {2008})}\BibitemShut {NoStop}%
\bibitem [{\citenamefont {Denissenko}\ \emph {et~al.}(2012)\citenamefont
  {Denissenko}, \citenamefont {Kantsler}, \citenamefont {Smith},\ and\
  \citenamefont {Kirkman-Brown}}]{denissenko2012}%
  \BibitemOpen
  \bibfield  {author} {\bibinfo {author} {\bibfnamefont {P.}~\bibnamefont
  {Denissenko}}, \bibinfo {author} {\bibfnamefont {V.}~\bibnamefont
  {Kantsler}}, \bibinfo {author} {\bibfnamefont {D.~J.}\ \bibnamefont
  {Smith}},\ and\ \bibinfo {author} {\bibfnamefont {J.}~\bibnamefont
  {Kirkman-Brown}},\ }\bibfield  {title} {\bibinfo {title} {Human spermatozoa
  migration in microchannels reveals boundary-following navigation},\
  }\href@noop {} {\bibfield  {journal} {\bibinfo  {journal} {Proc.\ Natl.\
  Acad.\ Sci.\ USA}\ }\textbf {\bibinfo {volume} {109}},\ \bibinfo {pages}
  {8007} (\bibinfo {year} {2012})}\BibitemShut {NoStop}%
\bibitem [{\citenamefont {Tung}\ \emph {et~al.}(2014)\citenamefont {Tung},
  \citenamefont {Ardon}, \citenamefont {Fiore}, \citenamefont {Suarez},\ and\
  \citenamefont {Wu}}]{tung2014}%
  \BibitemOpen
  \bibfield  {author} {\bibinfo {author} {\bibfnamefont {C.-K.}\ \bibnamefont
  {Tung}}, \bibinfo {author} {\bibfnamefont {F.}~\bibnamefont {Ardon}},
  \bibinfo {author} {\bibfnamefont {A.~G.}\ \bibnamefont {Fiore}}, \bibinfo
  {author} {\bibfnamefont {S.~S.}\ \bibnamefont {Suarez}},\ and\ \bibinfo
  {author} {\bibfnamefont {M.}~\bibnamefont {Wu}},\ }\bibfield  {title}
  {\bibinfo {title} {Cooperative roles of biological flow and surface
  topography in guiding sperm migration revealed by a microfluidic model},\
  }\href@noop {} {\bibfield  {journal} {\bibinfo  {journal} {Lab Chip}\
  }\textbf {\bibinfo {volume} {14}},\ \bibinfo {pages} {1348} (\bibinfo {year}
  {2014})}\BibitemShut {NoStop}%
\bibitem [{\citenamefont {Sipos}\ \emph {et~al.}(2015)\citenamefont {Sipos},
  \citenamefont {Nagy}, \citenamefont {Leonardo},\ and\ \citenamefont
  {Galajda}}]{sipos2015}%
  \BibitemOpen
  \bibfield  {author} {\bibinfo {author} {\bibfnamefont {O.}~\bibnamefont
  {Sipos}}, \bibinfo {author} {\bibfnamefont {K.}~\bibnamefont {Nagy}},
  \bibinfo {author} {\bibfnamefont {R.~D.}\ \bibnamefont {Leonardo}},\ and\
  \bibinfo {author} {\bibfnamefont {P.}~\bibnamefont {Galajda}},\ }\bibfield
  {title} {\bibinfo {title} {Hydrodynamic trapping of swimming bacteria by
  convex walls},\ }\href@noop {} {\bibfield  {journal} {\bibinfo  {journal}
  {Phys\ Rev.\ Lett.}\ }\textbf {\bibinfo {volume} {114}},\ \bibinfo {pages}
  {258104} (\bibinfo {year} {2015})}\BibitemShut {NoStop}%
\bibitem [{\citenamefont {{H. Shum and E. A. Gaffney}}(2015)}]{shum2015}%
  \BibitemOpen
  \bibfield  {author} {\bibinfo {author} {\bibnamefont {{H. Shum and E. A.
  Gaffney}}},\ }\bibfield  {title} {\bibinfo {title} {Hydrodynamic analysis of
  flagellated bacteria swimming in corners of rectangular channels},\
  }\href@noop {} {\bibfield  {journal} {\bibinfo  {journal} {Phys\ Rev.\ E}\
  }\textbf {\bibinfo {volume} {92}},\ \bibinfo {pages} {063016} (\bibinfo
  {year} {2015})}\BibitemShut {NoStop}%
\bibitem [{\citenamefont {Ishimoto}\ \emph {et~al.}(2016)\citenamefont
  {Ishimoto}, \citenamefont {Cosson},\ and\ \citenamefont
  {Gaffney}}]{ishimoto2016}%
  \BibitemOpen
  \bibfield  {author} {\bibinfo {author} {\bibfnamefont {K.}~\bibnamefont
  {Ishimoto}}, \bibinfo {author} {\bibfnamefont {J.}~\bibnamefont {Cosson}},\
  and\ \bibinfo {author} {\bibfnamefont {E.~A.}\ \bibnamefont {Gaffney}},\
  }\bibfield  {title} {\bibinfo {title} {A simulation study of sperm motility
  hydrodynamics near fish eggs and spheres},\ }\href@noop {} {\bibfield
  {journal} {\bibinfo  {journal} {J.\ Theor.\ Biol.}\ }\textbf {\bibinfo
  {volume} {389}},\ \bibinfo {pages} {187} (\bibinfo {year}
  {2016})}\BibitemShut {NoStop}%
\bibitem [{\citenamefont {Nosrati}\ \emph {et~al.}(2016)\citenamefont
  {Nosrati}, \citenamefont {Graham}, \citenamefont {Liu},\ and\ \citenamefont
  {Sinton}}]{nosrati2016}%
  \BibitemOpen
  \bibfield  {author} {\bibinfo {author} {\bibfnamefont {R.}~\bibnamefont
  {Nosrati}}, \bibinfo {author} {\bibfnamefont {P.~J.}\ \bibnamefont {Graham}},
  \bibinfo {author} {\bibfnamefont {Q.}~\bibnamefont {Liu}},\ and\ \bibinfo
  {author} {\bibfnamefont {D.}~\bibnamefont {Sinton}},\ }\bibfield  {title}
  {\bibinfo {title} {Predominance of sperm motion in corners},\ }\href@noop {}
  {\bibfield  {journal} {\bibinfo  {journal} {Sci.\ Rep.}\ }\textbf {\bibinfo
  {volume} {6}},\ \bibinfo {pages} {26669} (\bibinfo {year}
  {2016})}\BibitemShut {NoStop}%
\bibitem [{\citenamefont {Ostapenko}\ \emph {et~al.}(2018)\citenamefont
  {Ostapenko}, \citenamefont {Schwarzendahl}, \citenamefont {B\"oddeker},
  \citenamefont {Kreis}, \citenamefont {Cammann}, \citenamefont {Mazza},\ and\
  \citenamefont {Ba\"umchen}}]{ostapenko2018}%
  \BibitemOpen
  \bibfield  {author} {\bibinfo {author} {\bibfnamefont {T.}~\bibnamefont
  {Ostapenko}}, \bibinfo {author} {\bibfnamefont {F.~J.}\ \bibnamefont
  {Schwarzendahl}}, \bibinfo {author} {\bibfnamefont {T.~J.}\ \bibnamefont
  {B\"oddeker}}, \bibinfo {author} {\bibfnamefont {C.~T.}\ \bibnamefont
  {Kreis}}, \bibinfo {author} {\bibfnamefont {J.}~\bibnamefont {Cammann}},
  \bibinfo {author} {\bibfnamefont {M.~G.}\ \bibnamefont {Mazza}},\ and\
  \bibinfo {author} {\bibfnamefont {O.}~\bibnamefont {Ba\"umchen}},\ }\bibfield
   {title} {\bibinfo {title} {Curvature-guided motility of microalgae in
  geometric confinement},\ }\href@noop {} {\bibfield  {journal} {\bibinfo
  {journal} {Phys.\ Rev.\ Lett.}\ }\textbf {\bibinfo {volume} {120}},\ \bibinfo
  {pages} {068002} (\bibinfo {year} {2018})}\BibitemShut {NoStop}%
\bibitem [{\citenamefont {Nishiguchi}\ \emph {et~al.}(2018)\citenamefont
  {Nishiguchi}, \citenamefont {Aranson}, \citenamefont {Snezhko},\ and\
  \citenamefont {Sokolov}}]{nishiguchi2018}%
  \BibitemOpen
  \bibfield  {author} {\bibinfo {author} {\bibfnamefont {D.}~\bibnamefont
  {Nishiguchi}}, \bibinfo {author} {\bibfnamefont {I.~S.}\ \bibnamefont
  {Aranson}}, \bibinfo {author} {\bibfnamefont {A.}~\bibnamefont {Snezhko}},\
  and\ \bibinfo {author} {\bibfnamefont {A.}~\bibnamefont {Sokolov}},\
  }\bibfield  {title} {\bibinfo {title} {Engineering bacterial vortex lattice
  via direct laser lithography},\ }\href@noop {} {\bibfield  {journal}
  {\bibinfo  {journal} {Nat.\ Commun.}\ }\textbf {\bibinfo {volume} {9}},\
  \bibinfo {pages} {4486} (\bibinfo {year} {2018})}\BibitemShut {NoStop}%
\bibitem [{\citenamefont {Rode}\ \emph {et~al.}(2019)\citenamefont {Rode},
  \citenamefont {Elgeti},\ and\ \citenamefont {Gompper}}]{rode2019}%
  \BibitemOpen
  \bibfield  {author} {\bibinfo {author} {\bibfnamefont {S.}~\bibnamefont
  {Rode}}, \bibinfo {author} {\bibfnamefont {J.}~\bibnamefont {Elgeti}},\ and\
  \bibinfo {author} {\bibfnamefont {G.}~\bibnamefont {Gompper}},\ }\bibfield
  {title} {\bibinfo {title} {Sperm motility in modulated microchannels},\
  }\href@noop {} {\bibfield  {journal} {\bibinfo  {journal} {New\ J.\ Phys.}\
  }\textbf {\bibinfo {volume} {21}},\ \bibinfo {pages} {013016} (\bibinfo
  {year} {2019})}\BibitemShut {NoStop}%
\bibitem [{\citenamefont {Yang}\ \emph {et~al.}(2019)\citenamefont {Yang},
  \citenamefont {Shimogonya},\ and\ \citenamefont {Ishikawa}}]{yang2019}%
  \BibitemOpen
  \bibfield  {author} {\bibinfo {author} {\bibfnamefont {J.}~\bibnamefont
  {Yang}}, \bibinfo {author} {\bibfnamefont {Y.}~\bibnamefont {Shimogonya}},\
  and\ \bibinfo {author} {\bibfnamefont {T.}~\bibnamefont {Ishikawa}},\
  }\bibfield  {title} {\bibinfo {title} {Bacterial detachment from a wall with
  a bump line},\ }\href@noop {} {\bibfield  {journal} {\bibinfo  {journal}
  {Phys. Rev. E}\ }\textbf {\bibinfo {volume} {99}},\ \bibinfo {pages} {023104}
  (\bibinfo {year} {2019})}\BibitemShut {NoStop}%
\bibitem [{\citenamefont {Takagi}\ \emph {et~al.}(2014)\citenamefont {Takagi},
  \citenamefont {Palacci}, \citenamefont {Braunschweig}, \citenamefont
  {Shelley},\ and\ \citenamefont {Zhang}}]{takagi2014}%
  \BibitemOpen
  \bibfield  {author} {\bibinfo {author} {\bibfnamefont {D.}~\bibnamefont
  {Takagi}}, \bibinfo {author} {\bibfnamefont {J.}~\bibnamefont {Palacci}},
  \bibinfo {author} {\bibfnamefont {A.~B.}\ \bibnamefont {Braunschweig}},
  \bibinfo {author} {\bibfnamefont {M.~J.}\ \bibnamefont {Shelley}},\ and\
  \bibinfo {author} {\bibfnamefont {J.}~\bibnamefont {Zhang}},\ }\bibfield
  {title} {\bibinfo {title} {Hydrodynamic capture of microswimmers into
  sphere-bound orbits},\ }\href@noop {} {\bibfield  {journal} {\bibinfo
  {journal} {Soft Matter}\ }\textbf {\bibinfo {volume} {10}},\ \bibinfo {pages}
  {1784} (\bibinfo {year} {2014})}\BibitemShut {NoStop}%
\bibitem [{\citenamefont {Spagnolie}\ \emph {et~al.}(2015)\citenamefont
  {Spagnolie}, \citenamefont {Moreno-Flores}, \citenamefont {Bartolo},\ and\
  \citenamefont {Lauga}}]{spagnolie2015}%
  \BibitemOpen
  \bibfield  {author} {\bibinfo {author} {\bibfnamefont {S.~E.}\ \bibnamefont
  {Spagnolie}}, \bibinfo {author} {\bibfnamefont {G.~R.}\ \bibnamefont
  {Moreno-Flores}}, \bibinfo {author} {\bibfnamefont {D.}~\bibnamefont
  {Bartolo}},\ and\ \bibinfo {author} {\bibfnamefont {E.}~\bibnamefont
  {Lauga}},\ }\bibfield  {title} {\bibinfo {title} {Hgeometric capture and
  escape of a microswimmer colliding with an obstacle},\ }\href@noop {}
  {\bibfield  {journal} {\bibinfo  {journal} {Soft Matter}\ }\textbf {\bibinfo
  {volume} {11}},\ \bibinfo {pages} {3396} (\bibinfo {year}
  {2015})}\BibitemShut {NoStop}%
\bibitem [{\citenamefont {Liu}\ \emph {et~al.}(2016)\citenamefont {Liu},
  \citenamefont {Zhou}, \citenamefont {Wang},\ and\ \citenamefont
  {Zhang}}]{liu2016}%
  \BibitemOpen
  \bibfield  {author} {\bibinfo {author} {\bibfnamefont {C.}~\bibnamefont
  {Liu}}, \bibinfo {author} {\bibfnamefont {C.}~\bibnamefont {Zhou}}, \bibinfo
  {author} {\bibfnamefont {W.}~\bibnamefont {Wang}},\ and\ \bibinfo {author}
  {\bibfnamefont {H.~P.}\ \bibnamefont {Zhang}},\ }\bibfield  {title} {\bibinfo
  {title} {Bimetallic microswimmers speed up in confining channels},\
  }\href@noop {} {\bibfield  {journal} {\bibinfo  {journal} {Phys.\ Rev.\
  Lett.}\ }\textbf {\bibinfo {volume} {117}},\ \bibinfo {pages} {198001}
  (\bibinfo {year} {2016})}\BibitemShut {NoStop}%
\bibitem [{\citenamefont {Yang}\ \emph {et~al.}(2016)\citenamefont {Yang},
  \citenamefont {Qian}, \citenamefont {Zhao},\ and\ \citenamefont
  {Qiao}}]{yang2016}%
  \BibitemOpen
  \bibfield  {author} {\bibinfo {author} {\bibfnamefont {F.}~\bibnamefont
  {Yang}}, \bibinfo {author} {\bibfnamefont {S.}~\bibnamefont {Qian}}, \bibinfo
  {author} {\bibfnamefont {Y.}~\bibnamefont {Zhao}},\ and\ \bibinfo {author}
  {\bibfnamefont {R.}~\bibnamefont {Qiao}},\ }\bibfield  {title} {\bibinfo
  {title} {Self-diffusiophoresis of janus catalytic micromotors in confined
  geometries},\ }\href@noop {} {\bibfield  {journal} {\bibinfo  {journal}
  {Langmuir}\ }\textbf {\bibinfo {volume} {32}},\ \bibinfo {pages} {5580}
  (\bibinfo {year} {2016})}\BibitemShut {NoStop}%
\bibitem [{\citenamefont {Wykes}\ \emph {et~al.}(2017)\citenamefont {Wykes},
  \citenamefont {Zhong}, \citenamefont {J.Tong}, \citenamefont {Adachi},
  \citenamefont {Liu}, \citenamefont {Ristroph}, \citenamefont {Ward},
  \citenamefont {Shelley},\ and\ \citenamefont {Zhang}}]{wykes2017}%
  \BibitemOpen
  \bibfield  {author} {\bibinfo {author} {\bibfnamefont {M.~S.~D.}\
  \bibnamefont {Wykes}}, \bibinfo {author} {\bibfnamefont {X.}~\bibnamefont
  {Zhong}}, \bibinfo {author} {\bibnamefont {J.Tong}}, \bibinfo {author}
  {\bibfnamefont {T.}~\bibnamefont {Adachi}}, \bibinfo {author} {\bibfnamefont
  {Y.}~\bibnamefont {Liu}}, \bibinfo {author} {\bibfnamefont {L.}~\bibnamefont
  {Ristroph}}, \bibinfo {author} {\bibfnamefont {M.~D.}\ \bibnamefont {Ward}},
  \bibinfo {author} {\bibfnamefont {M.~J.}\ \bibnamefont {Shelley}},\ and\
  \bibinfo {author} {\bibfnamefont {J.}~\bibnamefont {Zhang}},\ }\bibfield
  {title} {\bibinfo {title} {Guiding microscale swimmers using teardrop-shaped
  posts},\ }\href@noop {} {\bibfield  {journal} {\bibinfo  {journal} {Soft
  Matter}\ }\textbf {\bibinfo {volume} {13}},\ \bibinfo {pages} {4681}
  (\bibinfo {year} {2017})}\BibitemShut {NoStop}%
\bibitem [{\citenamefont {{S. H. Rad and A. Najafi}}(2010)}]{rad2010}%
  \BibitemOpen
  \bibfield  {author} {\bibinfo {author} {\bibnamefont {{S. H. Rad and A.
  Najafi}}},\ }\bibfield  {title} {\bibinfo {title} {Hydrodynamic interactions
  of spherical particles in a fluid confined by a rough no-slip wall},\
  }\href@noop {} {\bibfield  {journal} {\bibinfo  {journal} {Phys.\ Rev.\ E}\
  }\textbf {\bibinfo {volume} {82}},\ \bibinfo {pages} {036305} (\bibinfo
  {year} {2010})}\BibitemShut {NoStop}%
\bibitem [{\citenamefont {Assoudi}\ \emph {et~al.}(2018)\citenamefont
  {Assoudi}, \citenamefont {Chaoui}, \citenamefont {Feuillebois},\ and\
  \citenamefont {Allouche}}]{assoudi2018}%
  \BibitemOpen
  \bibfield  {author} {\bibinfo {author} {\bibfnamefont {R.}~\bibnamefont
  {Assoudi}}, \bibinfo {author} {\bibfnamefont {M.}~\bibnamefont {Chaoui}},
  \bibinfo {author} {\bibfnamefont {F.}~\bibnamefont {Feuillebois}},\ and\
  \bibinfo {author} {\bibfnamefont {H.}~\bibnamefont {Allouche}},\ }\bibfield
  {title} {\bibinfo {title} {Motion of a spherical particle along a rough wall
  in a shear flow},\ }\href@noop {} {\bibfield  {journal} {\bibinfo  {journal}
  {Z.\ Angew.\ Math.\ Phys.}\ }\textbf {\bibinfo {volume} {69}},\ \bibinfo
  {pages} {112} (\bibinfo {year} {2018})}\BibitemShut {NoStop}%
\bibitem [{\citenamefont {{V. A. Shaik and A. M. Ardekani}}(2017)}]{shaik2017}%
  \BibitemOpen
  \bibfield  {author} {\bibinfo {author} {\bibnamefont {{V. A. Shaik and A. M.
  Ardekani}}},\ }\bibfield  {title} {\bibinfo {title} {Motion of a model
  swimmer near a weakly deforming interface},\ }\href@noop {} {\bibfield
  {journal} {\bibinfo  {journal} {J.\ Fluid Mech.}\ }\textbf {\bibinfo {volume}
  {824}},\ \bibinfo {pages} {42} (\bibinfo {year} {2017})}\BibitemShut
  {NoStop}%
\bibitem [{\citenamefont {Lighthill}(1952)}]{lighthill1952}%
  \BibitemOpen
  \bibfield  {author} {\bibinfo {author} {\bibfnamefont {M.~J.}\ \bibnamefont
  {Lighthill}},\ }\bibfield  {title} {\bibinfo {title} {On the squirming motion
  of nearly spherical deformable bodies through liquids at very small
  {R}eynolds numbers},\ }\href@noop {} {\bibfield  {journal} {\bibinfo
  {journal} {Commun.\ Pure Appl.\ Math.}\ }\textbf {\bibinfo {volume} {5}},\
  \bibinfo {pages} {109} (\bibinfo {year} {1952})}\BibitemShut {NoStop}%
\bibitem [{\citenamefont {Blake}(1971)}]{blake1971a}%
  \BibitemOpen
  \bibfield  {author} {\bibinfo {author} {\bibfnamefont {J.~R.}\ \bibnamefont
  {Blake}},\ }\bibfield  {title} {\bibinfo {title} {A spherical envelope
  approach to ciliary propulsion},\ }\href@noop {} {\bibfield  {journal}
  {\bibinfo  {journal} {J.\ Fluid Mech.}\ }\textbf {\bibinfo {volume} {46}},\
  \bibinfo {pages} {199} (\bibinfo {year} {1971})}\BibitemShut {NoStop}%
\bibitem [{\citenamefont {Pedley}(2016)}]{pedley2016}%
  \BibitemOpen
  \bibfield  {author} {\bibinfo {author} {\bibfnamefont {T.~J.}\ \bibnamefont
  {Pedley}},\ }\bibfield  {title} {\bibinfo {title} {Spherical squirmers:
  models for swimming micro-organisms},\ }\href@noop {} {\bibfield  {journal}
  {\bibinfo  {journal} {IMA J.\ Appl.\ Math.}\ }\textbf {\bibinfo {volume}
  {81}},\ \bibinfo {pages} {488} (\bibinfo {year} {2016})}\BibitemShut
  {NoStop}%
\bibitem [{\citenamefont {Pedley}\ \emph {et~al.}(2016)\citenamefont {Pedley},
  \citenamefont {Brumley},\ and\ \citenamefont {Goldstein}}]{pedley2016a}%
  \BibitemOpen
  \bibfield  {author} {\bibinfo {author} {\bibfnamefont {T.~J.}\ \bibnamefont
  {Pedley}}, \bibinfo {author} {\bibfnamefont {D.~R.}\ \bibnamefont
  {Brumley}},\ and\ \bibinfo {author} {\bibfnamefont {R.~E.}\ \bibnamefont
  {Goldstein}},\ }\bibfield  {title} {\bibinfo {title} {Squirmers with swirl: a
  model for volvox swimming},\ }\href@noop {} {\bibfield  {journal} {\bibinfo
  {journal} {J.\ Fluid Mech.}\ }\textbf {\bibinfo {volume} {798}},\ \bibinfo
  {pages} {165} (\bibinfo {year} {2016})}\BibitemShut {NoStop}%
\bibitem [{\citenamefont {Magar}\ \emph {et~al.}(2003)\citenamefont {Magar},
  \citenamefont {Goto},\ and\ \citenamefont {Pedley}}]{magar2003}%
  \BibitemOpen
  \bibfield  {author} {\bibinfo {author} {\bibfnamefont {V.}~\bibnamefont
  {Magar}}, \bibinfo {author} {\bibfnamefont {T.}~\bibnamefont {Goto}},\ and\
  \bibinfo {author} {\bibfnamefont {T.~J.}\ \bibnamefont {Pedley}},\ }\bibfield
   {title} {\bibinfo {title} {Nutrient uptake by a self propelled spherical
  squirmer},\ }\href@noop {} {\bibfield  {journal} {\bibinfo  {journal}
  {Q.~J.~Mech.~Appl.~Math}\ }\textbf {\bibinfo {volume} {56}},\ \bibinfo
  {pages} {65} (\bibinfo {year} {2003})}\BibitemShut {NoStop}%
\bibitem [{\citenamefont {Ishikawa}\ \emph {et~al.}(2006)\citenamefont
  {Ishikawa}, \citenamefont {Simmonds},\ and\ \citenamefont
  {Pedley}}]{ishikawa2006}%
  \BibitemOpen
  \bibfield  {author} {\bibinfo {author} {\bibfnamefont {T.}~\bibnamefont
  {Ishikawa}}, \bibinfo {author} {\bibfnamefont {M.~P.}\ \bibnamefont
  {Simmonds}},\ and\ \bibinfo {author} {\bibfnamefont {T.~J.}\ \bibnamefont
  {Pedley}},\ }\bibfield  {title} {\bibinfo {title} {Hydrodynamic interaction
  of two swimming model micro-organisms},\ }\href@noop {} {\bibfield  {journal}
  {\bibinfo  {journal} {J.\ Fluid Mech.}\ }\textbf {\bibinfo {volume} {568}},\
  \bibinfo {pages} {119} (\bibinfo {year} {2006})}\BibitemShut {NoStop}%
\bibitem [{\citenamefont {Drescher}\ \emph {et~al.}(2009)\citenamefont
  {Drescher}, \citenamefont {Leptos}, \citenamefont {Tucval}, \citenamefont
  {Ishikawa}, \citenamefont {Pedley},\ and\ \citenamefont
  {Goldstein}}]{drescher2009}%
  \BibitemOpen
  \bibfield  {author} {\bibinfo {author} {\bibfnamefont {K.}~\bibnamefont
  {Drescher}}, \bibinfo {author} {\bibfnamefont {K.~C.}\ \bibnamefont
  {Leptos}}, \bibinfo {author} {\bibfnamefont {I.}~\bibnamefont {Tucval}},
  \bibinfo {author} {\bibfnamefont {T.}~\bibnamefont {Ishikawa}}, \bibinfo
  {author} {\bibfnamefont {T.~J.}\ \bibnamefont {Pedley}},\ and\ \bibinfo
  {author} {\bibfnamefont {R.~E.}\ \bibnamefont {Goldstein}},\ }\bibfield
  {title} {\bibinfo {title} {Dancing {\it volvox}: Hydrodynamic bound states of
  swimming algae},\ }\href@noop {} {\bibfield  {journal} {\bibinfo  {journal}
  {Phys.~Rev.~Lett.}\ }\textbf {\bibinfo {volume} {102}},\ \bibinfo {pages}
  {168101} (\bibinfo {year} {2009})}\BibitemShut {NoStop}%
\bibitem [{\citenamefont {{S. E. Spagnolie and E.
  Lauga}}(2012)}]{spagnolie2012}%
  \BibitemOpen
  \bibfield  {author} {\bibinfo {author} {\bibnamefont {{S. E. Spagnolie and E.
  Lauga}}},\ }\bibfield  {title} {\bibinfo {title} {Hydrodynamics of
  self-propulsion near a boundary: predictions and accuracy of far-field
  approximations},\ }\href@noop {} {\bibfield  {journal} {\bibinfo  {journal}
  {J.\ Fluid Mech.}\ }\textbf {\bibinfo {volume} {700}},\ \bibinfo {pages}
  {105} (\bibinfo {year} {2012})}\BibitemShut {NoStop}%
\bibitem [{\citenamefont {{K. Ishimoto and E. A.
  Gaffney}}(2013)}]{ishimoto2013}%
  \BibitemOpen
  \bibfield  {author} {\bibinfo {author} {\bibnamefont {{K. Ishimoto and E. A.
  Gaffney}}},\ }\bibfield  {title} {\bibinfo {title} {Squirmer dynamics near a
  boundary},\ }\href@noop {} {\bibfield  {journal} {\bibinfo  {journal} {Phys.\
  Rev.\ E}\ }\textbf {\bibinfo {volume} {88}},\ \bibinfo {pages} {062702}
  (\bibinfo {year} {2013})}\BibitemShut {NoStop}%
\bibitem [{\citenamefont {Evans}\ \emph {et~al.}(2011)\citenamefont {Evans},
  \citenamefont {Ishikawa}, \citenamefont {Yamaguchi},\ and\ \citenamefont
  {Lauga}}]{evans2011}%
  \BibitemOpen
  \bibfield  {author} {\bibinfo {author} {\bibfnamefont {A.~A.}\ \bibnamefont
  {Evans}}, \bibinfo {author} {\bibfnamefont {T.}~\bibnamefont {Ishikawa}},
  \bibinfo {author} {\bibfnamefont {T.}~\bibnamefont {Yamaguchi}},\ and\
  \bibinfo {author} {\bibfnamefont {E.}~\bibnamefont {Lauga}},\ }\bibfield
  {title} {\bibinfo {title} {Orientational order in concentrated suspensions of
  spherical microswimmers},\ }\href@noop {} {\bibfield  {journal} {\bibinfo
  {journal} {Phys.\ Fluids}\ }\textbf {\bibinfo {volume} {23}},\ \bibinfo
  {pages} {111702} (\bibinfo {year} {2011})}\BibitemShut {NoStop}%
\bibitem [{\citenamefont {{A. Z\"ottl and H. Stark}}(2012)}]{zottl2014}%
  \BibitemOpen
  \bibfield  {author} {\bibinfo {author} {\bibnamefont {{A. Z\"ottl and H.
  Stark}}},\ }\bibfield  {title} {\bibinfo {title} {Hydrodynamics determines
  collective motion and phase behavior of active colloids in
  quasi-two-dimensional confinement},\ }\href@noop {} {\bibfield  {journal}
  {\bibinfo  {journal} {Phys.~Rev.~Lett.}\ }\textbf {\bibinfo {volume} {108}},\
  \bibinfo {pages} {118101} (\bibinfo {year} {2012})}\BibitemShut {NoStop}%
\bibitem [{\citenamefont {Delfau}\ \emph {et~al.}(2016)\citenamefont {Delfau},
  \citenamefont {Molina},\ and\ \citenamefont {Sano}}]{delfau2016}%
  \BibitemOpen
  \bibfield  {author} {\bibinfo {author} {\bibfnamefont {J.~B.}\ \bibnamefont
  {Delfau}}, \bibinfo {author} {\bibfnamefont {J.}~\bibnamefont {Molina}},\
  and\ \bibinfo {author} {\bibfnamefont {M.}~\bibnamefont {Sano}},\ }\bibfield
  {title} {\bibinfo {title} {Collective behavior of strongly confined
  suspensions of squirmers},\ }\href@noop {} {\bibfield  {journal} {\bibinfo
  {journal} {EPL}\ }\textbf {\bibinfo {volume} {114}},\ \bibinfo {pages}
  {24001} (\bibinfo {year} {2016})}\BibitemShut {NoStop}%
\bibitem [{\citenamefont {Oyama}\ \emph {et~al.}(2016)\citenamefont {Oyama},
  \citenamefont {Molina},\ and\ \citenamefont {Yamamoto}}]{oyama2016}%
  \BibitemOpen
  \bibfield  {author} {\bibinfo {author} {\bibfnamefont {N.}~\bibnamefont
  {Oyama}}, \bibinfo {author} {\bibfnamefont {J.~J.}\ \bibnamefont {Molina}},\
  and\ \bibinfo {author} {\bibfnamefont {R.}~\bibnamefont {Yamamoto}},\
  }\bibfield  {title} {\bibinfo {title} {Purely hydrodynamic origin for
  swarming of swimming particles},\ }\href@noop {} {\bibfield  {journal}
  {\bibinfo  {journal} {Phys.~Rev.~E}\ }\textbf {\bibinfo {volume} {93}},\
  \bibinfo {pages} {043114} (\bibinfo {year} {2016})}\BibitemShut {NoStop}%
\bibitem [{\citenamefont {Lauga}(2009)}]{lauga2009a}%
  \BibitemOpen
  \bibfield  {author} {\bibinfo {author} {\bibfnamefont {E.}~\bibnamefont
  {Lauga}},\ }\bibfield  {title} {\bibinfo {title} {Life at high
  $\mbox{D}$eborah number},\ }\href@noop {} {\bibfield  {journal} {\bibinfo
  {journal} {EPL}\ }\textbf {\bibinfo {volume} {86}},\ \bibinfo {pages} {64001}
  (\bibinfo {year} {2009})}\BibitemShut {NoStop}%
\bibitem [{\citenamefont {Zhu}\ \emph {et~al.}(2012)\citenamefont {Zhu},
  \citenamefont {Lauga},\ and\ \citenamefont {Brandt}}]{zhu2012}%
  \BibitemOpen
  \bibfield  {author} {\bibinfo {author} {\bibfnamefont {L.}~\bibnamefont
  {Zhu}}, \bibinfo {author} {\bibfnamefont {E.}~\bibnamefont {Lauga}},\ and\
  \bibinfo {author} {\bibfnamefont {L.}~\bibnamefont {Brandt}},\ }\bibfield
  {title} {\bibinfo {title} {Self-propulsion in viscoelastic fluids: Pushers
  vs. pullers},\ }\href@noop {} {\bibfield  {journal} {\bibinfo  {journal}
  {Phys.~Fluids}\ }\textbf {\bibinfo {volume} {24}},\ \bibinfo {pages} {051902}
  (\bibinfo {year} {2012})}\BibitemShut {NoStop}%
\bibitem [{\citenamefont {{H. Nganguia and O. S. Pak}}(2018)}]{nganguia2018}%
  \BibitemOpen
  \bibfield  {author} {\bibinfo {author} {\bibnamefont {{H. Nganguia and O. S.
  Pak}}},\ }\bibfield  {title} {\bibinfo {title} {Squirming motion in a
  $\mbox{B}$rinkman medium},\ }\href@noop {} {\bibfield  {journal} {\bibinfo
  {journal} {J.\ Fluid Mech.}\ }\textbf {\bibinfo {volume} {855}},\ \bibinfo
  {pages} {554} (\bibinfo {year} {2018})}\BibitemShut {NoStop}%
\bibitem [{\citenamefont {{I. Llopis and I.
  Pagonabarraga}}(2010)}]{llopis2010}%
  \BibitemOpen
  \bibfield  {author} {\bibinfo {author} {\bibnamefont {{I. Llopis and I.
  Pagonabarraga}}},\ }\bibfield  {title} {\bibinfo {title} {Hydrodynamic
  interactions in squirmer motion: Swimming with a neighbour and close to a
  wall},\ }\href@noop {} {\bibfield  {journal} {\bibinfo  {journal}
  {J.~Non-Newt.~Fluid Mech.}\ }\textbf {\bibinfo {volume} {165}},\ \bibinfo
  {pages} {946} (\bibinfo {year} {2010})}\BibitemShut {NoStop}%
\bibitem [{\citenamefont {Zhu}\ \emph {et~al.}(2013)\citenamefont {Zhu},
  \citenamefont {Lauga},\ and\ \citenamefont {Brandt}}]{zhu2013}%
  \BibitemOpen
  \bibfield  {author} {\bibinfo {author} {\bibfnamefont {L.}~\bibnamefont
  {Zhu}}, \bibinfo {author} {\bibfnamefont {E.}~\bibnamefont {Lauga}},\ and\
  \bibinfo {author} {\bibfnamefont {L.}~\bibnamefont {Brandt}},\ }\bibfield
  {title} {\bibinfo {title} {Low-$\mbox{R}$eynolds-number swimmer in a
  capillary tube},\ }\href@noop {} {\bibfield  {journal} {\bibinfo  {journal}
  {J.\ Fluid Mech.}\ }\textbf {\bibinfo {volume} {725}},\ \bibinfo {pages}
  {285} (\bibinfo {year} {2013})}\BibitemShut {NoStop}%
\bibitem [{\citenamefont {Chamolly}\ \emph {et~al.}(2017)\citenamefont
  {Chamolly}, \citenamefont {Ishikawa},\ and\ \citenamefont
  {Lauga}}]{chamolly2017}%
  \BibitemOpen
  \bibfield  {author} {\bibinfo {author} {\bibfnamefont {A.}~\bibnamefont
  {Chamolly}}, \bibinfo {author} {\bibfnamefont {T.}~\bibnamefont {Ishikawa}},\
  and\ \bibinfo {author} {\bibfnamefont {E.}~\bibnamefont {Lauga}},\ }\bibfield
   {title} {\bibinfo {title} {Active particles in periodic lattices},\
  }\href@noop {} {\bibfield  {journal} {\bibinfo  {journal} {New J.\ Phys.}\
  }\textbf {\bibinfo {volume} {19}},\ \bibinfo {pages} {115001} (\bibinfo
  {year} {2017})}\BibitemShut {NoStop}%
\bibitem [{\citenamefont {{S. Das and A. Cacciuto}}(2019)}]{das2019}%
  \BibitemOpen
  \bibfield  {author} {\bibinfo {author} {\bibnamefont {{S. Das and A.
  Cacciuto}}},\ }\bibfield  {title} {\bibinfo {title} {Colloidal swimmers near
  curved and structured walls},\ }\href@noop {} {\bibfield  {journal} {\bibinfo
   {journal} {Soft Matter}\ }\textbf {\bibinfo {volume} {15}},\ \bibinfo
  {pages} {8290} (\bibinfo {year} {2019})}\BibitemShut {NoStop}%
\bibitem [{\citenamefont {Simmchen}\ \emph {et~al.}(2016)\citenamefont
  {Simmchen}, \citenamefont {Katuri}, \citenamefont {Uspal}, \citenamefont
  {Popescu}, \citenamefont {Tasinkevych},\ and\ \citenamefont
  {Sanchez}}]{simmchen2016}%
  \BibitemOpen
  \bibfield  {author} {\bibinfo {author} {\bibfnamefont {J.}~\bibnamefont
  {Simmchen}}, \bibinfo {author} {\bibfnamefont {J.}~\bibnamefont {Katuri}},
  \bibinfo {author} {\bibfnamefont {W.~E.}\ \bibnamefont {Uspal}}, \bibinfo
  {author} {\bibfnamefont {M.~N.}\ \bibnamefont {Popescu}}, \bibinfo {author}
  {\bibfnamefont {M.}~\bibnamefont {Tasinkevych}},\ and\ \bibinfo {author}
  {\bibfnamefont {S.}~\bibnamefont {Sanchez}},\ }\bibfield  {title} {\bibinfo
  {title} {Topographical pathways guide chemical microswimmers},\ }\href@noop
  {} {\bibfield  {journal} {\bibinfo  {journal} {Nat.\ Commun.}\ }\textbf
  {\bibinfo {volume} {7}},\ \bibinfo {pages} {10598} (\bibinfo {year}
  {2016})}\BibitemShut {NoStop}%
\bibitem [{\citenamefont {Klein}\ \emph {et~al.}(2003)\citenamefont {Klein},
  \citenamefont {Clapp},\ and\ \citenamefont {Dickinson}}]{dickinson2003}%
  \BibitemOpen
  \bibfield  {author} {\bibinfo {author} {\bibfnamefont {J.}~\bibnamefont
  {Klein}}, \bibinfo {author} {\bibfnamefont {A.}~\bibnamefont {Clapp}},\ and\
  \bibinfo {author} {\bibfnamefont {R.~B.}\ \bibnamefont {Dickinson}},\
  }\bibfield  {title} {\bibinfo {title} {Direct measurement of interaction
  forces between a single bacterium and a flat plate},\ }\href@noop {}
  {\bibfield  {journal} {\bibinfo  {journal} {Journal of Colloid and Interface
  Science}\ }\textbf {\bibinfo {volume} {261}},\ \bibinfo {pages} {379}
  (\bibinfo {year} {2003})}\BibitemShut {NoStop}%
\bibitem [{\citenamefont {Lintuvuori}\ \emph {et~al.}(2016)\citenamefont
  {Lintuvuori}, \citenamefont {Brown}, \citenamefont {Stratford},\ and\
  \citenamefont {Marenduzzo}}]{lintuvuori2016}%
  \BibitemOpen
  \bibfield  {author} {\bibinfo {author} {\bibfnamefont {J.~S.}\ \bibnamefont
  {Lintuvuori}}, \bibinfo {author} {\bibfnamefont {A.~T.}\ \bibnamefont
  {Brown}}, \bibinfo {author} {\bibfnamefont {K.}~\bibnamefont {Stratford}},\
  and\ \bibinfo {author} {\bibfnamefont {D.}~\bibnamefont {Marenduzzo}},\
  }\bibfield  {title} {\bibinfo {title} {Hydrodynamic oscillations and variable
  swimming speed in squirmers close to repulsive walls},\ }\href@noop {}
  {\bibfield  {journal} {\bibinfo  {journal} {Soft Matter}\ }\textbf {\bibinfo
  {volume} {12}},\ \bibinfo {pages} {7959} (\bibinfo {year}
  {2016})}\BibitemShut {NoStop}%
\bibitem [{\citenamefont {Ishimoto}(2017)}]{ishimoto2017}%
  \BibitemOpen
  \bibfield  {author} {\bibinfo {author} {\bibfnamefont {K.}~\bibnamefont
  {Ishimoto}},\ }\bibfield  {title} {\bibinfo {title} {Guidance of
  microswimmers by wall and flow: Thigmotaxis and rheotaxis of unsteady
  squirmers in two and three dimensions},\ }\href@noop {} {\bibfield  {journal}
  {\bibinfo  {journal} {Phys.\ Rev.\ E}\ }\textbf {\bibinfo {volume} {96}},\
  \bibinfo {pages} {043101} (\bibinfo {year} {2017})}\BibitemShut {NoStop}%
\bibitem [{\citenamefont {Walker}\ \emph {et~al.}(2019)\citenamefont {Walker},
  \citenamefont {Wheeler}, \citenamefont {Ishimoto},\ and\ \citenamefont
  {Gaffney}}]{walker2019}%
  \BibitemOpen
  \bibfield  {author} {\bibinfo {author} {\bibfnamefont {B.~J.}\ \bibnamefont
  {Walker}}, \bibinfo {author} {\bibfnamefont {R.~J.}\ \bibnamefont {Wheeler}},
  \bibinfo {author} {\bibfnamefont {K.}~\bibnamefont {Ishimoto}},\ and\
  \bibinfo {author} {\bibfnamefont {E.~A.}\ \bibnamefont {Gaffney}},\
  }\bibfield  {title} {\bibinfo {title} {Boundary behaviours of {\it
  $\mbox{l}$eishmania mexicana}: a hydrodynamic simulation study},\ }\href@noop
  {} {\bibfield  {journal} {\bibinfo  {journal} {J.\ Theor.\ Biol}\ }\textbf
  {\bibinfo {volume} {462}},\ \bibinfo {pages} {311} (\bibinfo {year}
  {2019})}\BibitemShut {NoStop}%
\bibitem [{\citenamefont {Uspal}\ \emph {et~al.}(2014)\citenamefont {Uspal},
  \citenamefont {Popescu}, \citenamefont {Dietrich},\ and\ \citenamefont
  {Tasinkevych}}]{uspal2015}%
  \BibitemOpen
  \bibfield  {author} {\bibinfo {author} {\bibfnamefont {W.~E.}\ \bibnamefont
  {Uspal}}, \bibinfo {author} {\bibfnamefont {M.~N.}\ \bibnamefont {Popescu}},
  \bibinfo {author} {\bibfnamefont {S.}~\bibnamefont {Dietrich}},\ and\
  \bibinfo {author} {\bibfnamefont {M.}~\bibnamefont {Tasinkevych}},\
  }\bibfield  {title} {\bibinfo {title} {Rheotaxis of spherical active
  particles near a planar wall},\ }\href@noop {} {\bibfield  {journal}
  {\bibinfo  {journal} {Soft Matter}\ }\textbf {\bibinfo {volume} {11}},\
  \bibinfo {pages} {6613} (\bibinfo {year} {2014})}\BibitemShut {NoStop}%
\bibitem [{\citenamefont {Smith}(2018)}]{smith2018}%
  \BibitemOpen
  \bibfield  {author} {\bibinfo {author} {\bibfnamefont {D.~J.}\ \bibnamefont
  {Smith}},\ }\bibfield  {title} {\bibinfo {title} {A nearest neighbor
  discretisation of the regularized $\mbox{S}$tokeslet boundary integral
  equation},\ }\href@noop {} {\bibfield  {journal} {\bibinfo  {journal} {J.\
  Comput.\ Phys.}\ }\textbf {\bibinfo {volume} {358}},\ \bibinfo {pages} {88}
  (\bibinfo {year} {2018})}\BibitemShut {NoStop}%
\bibitem [{\citenamefont {{M. T. Gallagher and D. J.
  Smith}}(2018)}]{gallagher2018}%
  \BibitemOpen
  \bibfield  {author} {\bibinfo {author} {\bibnamefont {{M. T. Gallagher and D.
  J. Smith}}},\ }\bibfield  {title} {\bibinfo {title} {Meshfree and efficient
  modeling of swimming cells},\ }\href@noop {} {\bibfield  {journal} {\bibinfo
  {journal} {Phys.\ Rev.\ Fluids}\ }\textbf {\bibinfo {volume} {3}},\ \bibinfo
  {pages} {053101} (\bibinfo {year} {2018})}\BibitemShut {NoStop}%
\bibitem [{\citenamefont {Pozrikidis}(1992)}]{pozrikidis1992}%
  \BibitemOpen
  \bibfield  {author} {\bibinfo {author} {\bibfnamefont {C.}~\bibnamefont
  {Pozrikidis}},\ }\href@noop {} {\emph {\bibinfo {title} {Boundary integral
  and singularity methods for linearized viscous flow}}}\ (\bibinfo
  {publisher} {Cambridge University Press},\ \bibinfo {year}
  {1992})\BibitemShut {NoStop}%
\bibitem [{\citenamefont {Cortez}(2001)}]{cortez2001}%
  \BibitemOpen
  \bibfield  {author} {\bibinfo {author} {\bibfnamefont {R.}~\bibnamefont
  {Cortez}},\ }\bibfield  {title} {\bibinfo {title} {The method of regularized
  $\mbox{S}$tokeslets},\ }\href@noop {} {\bibfield  {journal} {\bibinfo
  {journal} {SIAM J.\ Sci.\ Comput.}\ }\textbf {\bibinfo {volume} {23}},\
  \bibinfo {pages} {1204} (\bibinfo {year} {2001})}\BibitemShut {NoStop}%
\bibitem [{\citenamefont {Cortez}\ \emph {et~al.}(2005)\citenamefont {Cortez},
  \citenamefont {Fauci},\ and\ \citenamefont {Medovikov}}]{cortez2005}%
  \BibitemOpen
  \bibfield  {author} {\bibinfo {author} {\bibfnamefont {R.}~\bibnamefont
  {Cortez}}, \bibinfo {author} {\bibfnamefont {L.}~\bibnamefont {Fauci}},\ and\
  \bibinfo {author} {\bibfnamefont {A.}~\bibnamefont {Medovikov}},\ }\bibfield
  {title} {\bibinfo {title} {The method of regularized $\mbox{S}$tokeslets in
  three dimensions: Analysis, validation, and application to helical
  swimming},\ }\href@noop {} {\bibfield  {journal} {\bibinfo  {journal} {Phys.\
  Fluids}\ }\textbf {\bibinfo {volume} {17}},\ \bibinfo {pages} {031504}
  (\bibinfo {year} {2005})}\BibitemShut {NoStop}%
\bibitem [{\citenamefont {Ainley}\ \emph {et~al.}(2008)\citenamefont {Ainley},
  \citenamefont {Durkin}, \citenamefont {Embid}, \citenamefont {Boindala},\
  and\ \citenamefont {Cortez}}]{ainley2008}%
  \BibitemOpen
  \bibfield  {author} {\bibinfo {author} {\bibfnamefont {J.}~\bibnamefont
  {Ainley}}, \bibinfo {author} {\bibfnamefont {S.}~\bibnamefont {Durkin}},
  \bibinfo {author} {\bibfnamefont {R.}~\bibnamefont {Embid}}, \bibinfo
  {author} {\bibfnamefont {P.}~\bibnamefont {Boindala}},\ and\ \bibinfo
  {author} {\bibfnamefont {R.}~\bibnamefont {Cortez}},\ }\bibfield  {title}
  {\bibinfo {title} {The method of images for regularized
  $\mbox{S}$tokeslets},\ }\href@noop {} {\bibfield  {journal} {\bibinfo
  {journal} {J.\ Comput.\ Phys.}\ }\textbf {\bibinfo {volume} {4227}},\
  \bibinfo {pages} {4600} (\bibinfo {year} {2008})}\BibitemShut {NoStop}%
\bibitem [{\citenamefont {{K. Sarkar and A. Prosperetti}}(1996)}]{sarkar1996}%
  \BibitemOpen
  \bibfield  {author} {\bibinfo {author} {\bibnamefont {{K. Sarkar and A.
  Prosperetti}}},\ }\bibfield  {title} {\bibinfo {title} {Effective boundary
  conditions for $\mbox{S}$tokes flow over a rough surface},\ }\href@noop {}
  {\bibfield  {journal} {\bibinfo  {journal} {J.\ Fluid Mech.}\ }\textbf
  {\bibinfo {volume} {316}},\ \bibinfo {pages} {223} (\bibinfo {year}
  {1996})}\BibitemShut {NoStop}%
\bibitem [{\citenamefont {Kunert}\ \emph {et~al.}(2010)\citenamefont {Kunert},
  \citenamefont {Harting},\ and\ \citenamefont {Vinogradova}}]{kunert2010}%
  \BibitemOpen
  \bibfield  {author} {\bibinfo {author} {\bibfnamefont {C.}~\bibnamefont
  {Kunert}}, \bibinfo {author} {\bibfnamefont {J.}~\bibnamefont {Harting}},\
  and\ \bibinfo {author} {\bibfnamefont {O.~I.}\ \bibnamefont {Vinogradova}},\
  }\bibfield  {title} {\bibinfo {title} {Random-roughness hydrodynamic boundary
  conditions},\ }\href@noop {} {\bibfield  {journal} {\bibinfo  {journal}
  {Phys.\ Rev.\ Lett.}\ }\textbf {\bibinfo {volume} {105}},\ \bibinfo {pages}
  {016001} (\bibinfo {year} {2010})}\BibitemShut {NoStop}%
\bibitem [{\citenamefont {Luchini}(2013)}]{luchini2013}%
  \BibitemOpen
  \bibfield  {author} {\bibinfo {author} {\bibfnamefont {P.}~\bibnamefont
  {Luchini}},\ }\bibfield  {title} {\bibinfo {title} {Linearised no-slip
  boundary conditions at a rough surface},\ }\href@noop {} {\bibfield
  {journal} {\bibinfo  {journal} {J. Fluid Mech.}\ }\textbf {\bibinfo {volume}
  {737}},\ \bibinfo {pages} {349} (\bibinfo {year} {2013})}\BibitemShut
  {NoStop}%
\bibitem [{\citenamefont {{K. Ishimoto and E. A.
  Gaffney}}(2014)}]{ishimoto2014}%
  \BibitemOpen
  \bibfield  {author} {\bibinfo {author} {\bibnamefont {{K. Ishimoto and E. A.
  Gaffney}}},\ }\bibfield  {title} {\bibinfo {title} {A study of spermatozoan
  swimming stability near a surface},\ }\href@noop {} {\bibfield  {journal}
  {\bibinfo  {journal} {J.\ Theor.\ Biol}\ }\textbf {\bibinfo {volume} {360}},\
  \bibinfo {pages} {187} (\bibinfo {year} {2014})}\BibitemShut {NoStop}%
\bibitem [{\citenamefont {Smith}\ \emph {et~al.}(2008)\citenamefont {Smith},
  \citenamefont {Gaffney},\ and\ \citenamefont {Blake}}]{smith2008}%
  \BibitemOpen
  \bibfield  {author} {\bibinfo {author} {\bibfnamefont {D.~J.}\ \bibnamefont
  {Smith}}, \bibinfo {author} {\bibfnamefont {E.~A.}\ \bibnamefont {Gaffney}},\
  and\ \bibinfo {author} {\bibfnamefont {J.~R.}\ \bibnamefont {Blake}},\
  }\bibfield  {title} {\bibinfo {title} {Modelling mucociliary clearance},\
  }\href@noop {} {\bibfield  {journal} {\bibinfo  {journal}
  {Respir.~Physiol.~Neurobiol.}\ }\textbf {\bibinfo {volume} {163}},\ \bibinfo
  {pages} {178} (\bibinfo {year} {2008})}\BibitemShut {NoStop}%
\bibitem [{\citenamefont {Das}\ \emph {et~al.}(2020)\citenamefont {Das},
  \citenamefont {Jalilvand}, \citenamefont {Popescu}, \citenamefont {Uspal},
  \citenamefont {Dietrich},\ and\ \citenamefont {Kretzschmar}}]{das2020}%
  \BibitemOpen
  \bibfield  {author} {\bibinfo {author} {\bibfnamefont {S.}~\bibnamefont
  {Das}}, \bibinfo {author} {\bibfnamefont {Z.}~\bibnamefont {Jalilvand}},
  \bibinfo {author} {\bibfnamefont {M.~N.}\ \bibnamefont {Popescu}}, \bibinfo
  {author} {\bibfnamefont {W.~E.}\ \bibnamefont {Uspal}}, \bibinfo {author}
  {\bibfnamefont {S.}~\bibnamefont {Dietrich}},\ and\ \bibinfo {author}
  {\bibfnamefont {I.}~\bibnamefont {Kretzschmar}},\ }\bibfield  {title}
  {\bibinfo {title} {Floor- or ceiling-sliding for chemically active,
  gyrotactic, sedimenting janus particles},\ }\href@noop {} {\bibfield
  {journal} {\bibinfo  {journal} {Langmuir}\ }\textbf {\bibinfo {volume}
  {36}},\ \bibinfo {pages} {7133} (\bibinfo {year} {2020})}\BibitemShut
  {NoStop}%
\bibitem [{\citenamefont {{M. T. Gallagher and D. J.
  Smith}}(2020)}]{gallagher2020}%
  \BibitemOpen
  \bibfield  {author} {\bibinfo {author} {\bibnamefont {{M. T. Gallagher and D.
  J. Smith}}},\ }\bibfield  {title} {\bibinfo {title} {Passively parallel
  regularized stokeslets},\ }\href@noop {} {\bibfield  {journal} {\bibinfo
  {journal} {Phil.\ Trans.\ R.\ Soc.\ A}\ }\textbf {\bibinfo {volume} {378}},\
  \bibinfo {pages} {20190528} (\bibinfo {year} {2020})}\BibitemShut {NoStop}%
\end{thebibliography}%

\end{document}